\documentclass[fleqn,usenatbib]{mnras}

\usepackage{newtxtext,newtxmath}
\usepackage[T1]{fontenc}

\DeclareRobustCommand{\VAN}[3]{#2}
\let\VANthebibliography\thebibliography
\def\thebibliography{\DeclareRobustCommand{\VAN}[3]{##3}\VANthebibliography}

\usepackage{graphicx}	% Including figure files
\usepackage{amsmath}	% Advanced maths commands
\title[Wedge-recovery and high-$z$ galaxy mapping]{Machine-learning recovery of foreground wedge-removed 21-cm light cones for high-$z$ galaxy mapping}

\author[Kennedy et al.]{Jacob Kennedy$^{1}$\thanks{jacob.kennedy@mail.mcgill.ca}, Jonathan Colaço Carr$^{2}$, Samuel Gagnon-Hartman$^{1,3}$, Adrian Liu$^{1}$, Jordan Mirocha$^{1,4,5}$,
\newauthor{Yue Cui$^{6}$}
\\
$^{1}$Department of Physics and Trottier Space Institute, McGill University, Montreal, QC, Canada H3A 2T8 \\
$^{2}$School of Computer Science, McGill University, 845 Sherbrooke Street, Montreal H3A 0G4, Canada \\
$^{3}$Scuola Normale Superiore, Piazza dei Cavalleri 7, 56126 Pisa, Italy \\
$^{4}$Jet Propulsion Laboratory, California Institute of Technology, 4800 Oak Grove Drive, Pasadena, CA 91109, USA \\
$^{5}$California Institute of Technology, 1200 E. California Boulevard, Pasadena, CA 91125, USA \\
$^{6}$Department of Computer Science and Engineering, The Hong Kong University of Science and Technology, Clear Water Bay, Hong Kong SAR
}

\pubyear{2023}

\begin{document}
\label{firstpage}
\pagerange{\pageref{firstpage}--\pageref{lastpage}}
\maketitle

\begin{abstract}
\noindent Upcoming experiments will map the spatial distribution of the 21-cm signal over three-dimensional volumes of space during the Epoch of Reionization (EoR). Several methods have been proposed to mitigate the issue of astrophysical foreground contamination in tomographic images of the 21-cm signal, one of which involves the excision of a wedge-shaped region in cylindrical Fourier space. While this removes the $k$-modes most readily contaminated by foregrounds, the concurrent removal of cosmological information located within the wedge considerably distorts the structure of 21-cm images. In this study, we build upon a U-Net based deep learning algorithm to reconstruct foreground wedge-removed maps of the 21-cm signal, newly incorporating light-cone effects. Adopting the Square Kilometre Array (SKA) as our fiducial instrument, we highlight that our U-Net recovery framework retains a reasonable level of reliability even in the face of instrumental limitations and noise. We subsequently evaluate the efficacy of recovered maps in guiding high-redshift galaxy searches and providing context to existing galaxy catalogues. This will allow for studies of how the high-redshift galaxy luminosity function varies across environments, and ultimately refine our understanding of the connection between the ionization state of the intergalactic medium (IGM) and galaxies during the EoR.

\end{abstract}

\begin{keywords}
methods: data analysis -- dark ages, reionization, first stars -- cosmology: observations -- galaxies: high-redshift.
\end{keywords}

%%%%%%%%%%%%%%%%%%%%%%%%%%%%%%%%%%%%%%%%%%%%%%%%%%

%%%%%%%%%%%%%%%%% BODY OF PAPER %%%%%%%%%%%%%%%%%%

\section{Introduction}

The redshifted 21-cm spectral line has been identified as a promising probe of the early Universe and cosmic structure formation (e.g. \citealt{FURLANETTO2006181, morales2010_review, Pritchard_2012, Liu2020}). Ongoing and future experiments (LOw Frequency ARray (LOFAR); \citealt{LOFAR}, Murchison Widefield Array (MWA); \citealt{MWA}, Square Kilometre Array (SKA); \citealt{SKA2015}, Hydrogen Epoch of Reionization Array (HERA); \citealt{HERA2017}) will measure spatial fluctuations in the 21-cm signal from the Cosmic Dawn through the Epoch of Reionization (EoR). While several current experiments (LOFAR, MWA, HERA) have been optimally designed to measure the statistical properties of the 21-cm signal via the power spectrum, the upcoming SKA is configured to produce three-dimensional tomographic maps of the 21-cm signal over a wide range of redshifts. These maps will trace the evolution of the IGM's ionization state through the growth of ionized bubbles, or HII regions (e.g. \citealt{Morales_2010}). While it is expected that galaxies were the primary source of ionizing photons during the EoR (e.g. \citealt{Yan_2004_gals,Bouwens_2012_gals,Finkelstein_2012_gals,Robertson_2013_gals}), the properties of these source galaxies remain poorly constrained (e.g. \citealt{Robertson_2010}). Leveraging the sensitivity of the 21-cm background to the complete galaxy population presents the opportunity to indirectly constrain galaxy properties via their impact on the IGM. Therefore, by connecting 21-cm tomography to galaxy surveys, we may identify the galaxies responsible for reionization and ultimately improve our understanding of galaxy populations during this epoch. 

\begin{figure}
	\includegraphics[width=\columnwidth]{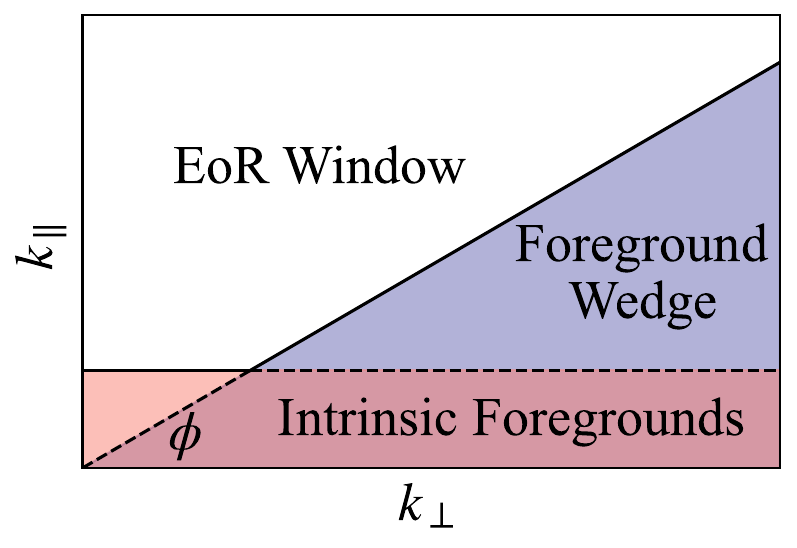}
    \caption{A qualitative picture of the footprint of foreground contamination relevant to 21-cm radio interferometers. While intrinsic foregrounds uniformly contaminate low $k_{\parallel}$ modes, foreground leakage beyond this region leads to the formation of the foreground wedge, parameterized by the wedge angle $\phi$. The EoR Window denotes the region of Fourier space where foregrounds are suppressed, in principle allowing for a clean measurement of the 21-cm signal.}
    \label{fig:foreground_wedge}
\end{figure}

The primary obstacle to 21-cm tomography is astrophysical foreground contamination. Galactic and extra-galactic foregrounds can be up to three to four orders of magnitude brighter than the cosmological signal (\citealt{Bernardi_2009, Bernardi_2010}). While foreground emission is expected to be spectrally smooth, further complications arise due to the interferometric nature of observing instruments. Various methods have been proposed to resolve the issue of foreground contamination in the context of the EoR 21-cm signal (see \citealt{Liu2020} for a summary). Several of these address the problem of mode-mixing, whereby the chromatic response of interferometers causes foregrounds to leak from lower to higher $k_{\parallel}$ modes (Fourier wavenumbers parallel to the line-of-sight; LoS). This process leads to the localization of foregrounds in a wedge-shaped region of Fourier space, known as the foreground wedge (\citealt{Datta_2010, Morales_2010, Parsons_2012, Vedantham_2012, Trott_2012, Hazelton_2013, Pober_2013, Thyagarajan_2013, Liu_2014a, Liu_2014b}). The boundary of the wedge can be expressed as a function of the cylindrical Fourier space coordinates ($k_{\perp}, k_{\parallel}$),
\begin{equation}
k_{\parallel} = k_{\perp}\frac{\text{sin}\theta_{\text{FoV}}E(z)}{(1+z)} \int_0^z \frac{dz'}{E(z')}\equiv k_{\perp} \text{tan}\phi
\label{eq:wedge}
\end{equation}

\noindent where $E(z) \equiv \sqrt{\Omega_m (1+z)^3 + \Omega_{\Lambda}}$, $\theta_{\text{FoV}}$ is the angular radius of the field of view of the interferometer, $\Omega_m$ is the normalized matter density, and $\Omega_{\Lambda}$ is the normalized dark energy density. The equivalency in Equation \eqref{eq:wedge} defines $\phi$ to be the angle between the wedge boundary and the $k_{\perp}$-axis (Fourier modes perpendicular to the LoS). A visual depiction of the foreground wedge is presented in Figure \ref{fig:foreground_wedge}, illustrated in the two-dimensional plane of cylindrical Fourier space. While the complementary EoR Window in Figure \ref{fig:foreground_wedge} denotes the area of Fourier space that is not readily contaminated by foregrounds, restricting one's observations to this region ensures any cosmological information located within the wedge is not accessed. This method of foreground avoidance alone is thus unfavourable for 21-cm tomography given \textit{all} Fourier modes are necessary for high fidelity imaging (\citealt{Liu2020}). Such complications motivate the consideration of alternative foreground mitigation techniques.

Recently, deep learning-based approaches to foreground removal have been presented in the literature (e.g. \citealt{Li_2019, Makinen_2021, GagnonHartman2021, bianco2023deep}). In particular, \cite{GagnonHartman2021}, henceforth referred to as GH21, demonstrated the success of a U-Net-based deep learning algorithm to recover Fourier modes that are obscured by foregrounds. Their algorithm does not rely on any knowledge of the foregrounds themselves, and enables the reconstruction of mock 21-cm images at fixed redshifts after all modes lying within the foreground wedge have been completely nulled out. Importantly, GH21 showed that the reconstruction of ionized bubbles was still possible even in the presence of such an aggressive foreground filter. More recently, \cite{bianco2023deep} demonstrated an alternative U-Net based-architecture that was able to identify neutral and ionized regions in foreground contaminated 21-cm maps. The work of \cite{bianco2023deep} differs from GH21 however, in that \cite{bianco2023deep} simulated foreground removal with a wide variety of methods including Principal Component Analysis, foreground wedge removal, and polynomial fitting to simulate realistic foreground residuals. Importantly, both GH21 and \cite{bianco2023deep} noted the applicability of such reconstructive algorithms for follow-up observations of ionizing sources, notably galaxies, located within recovered ionized regions.

In this study, we advance the work of GH21 in two primary ways. First, we modify the U-Net algorithm employed to improve recovery performance post-foreground wedge excision, and extend the recovery effort to include 21-cm maps with light-cone effects (\citealt{Datta2014_lightcone, La_Plante_2014_lightcone}). Given light-cone effects will be implicit in future observations of the EoR, this effort will provide a more realistic picture of the success of the U-Net recovery methodology presented in GH21. As the second and principal advancement, we demonstrate how one can use U-Net recovered 21-cm light-cones to direct high-redshift galaxy surveys and provide context to existing galaxy catalogues. The basis for our investigation stems from the prospect of obtaining foreground-contaminated tomographic maps of the 21-cm signal from experiments such as SKA. Applying the foreground removal algorithm and subsequent U-Net reconstruction framework we detail in this paper, one can identify the location, morphology, and size of ionized bubbles in 21-cm light-cones. To study the connection between ionized bubbles and photon-producing galaxies, one can use foreground-decontaminated maps to guide searches for these galaxies and supply information regarding the ionization environment of existing galaxy observations.  

Because 21-cm intensity mapping surveys such as SKA are designed to image an expansive field with a relatively limited angular resolution, they are unable to resolve individual galaxies at EoR redshifts. The required angular resolution is, however, well within the capabilities of current and next-generation galaxy surveys such as the James Webb Space Telescope (JWST; \citealt{JWST_2006}) and the Nancy Grace Roman Space Telescope (\emph{Roman}; \citealt{WFIRST_2019}). Each of these instruments have a much higher resolution and smaller field-of-view compared to SKA, possessing sufficient sensitivity to identify and study individual galaxies during the EoR (\citealt{Beardsley_2015}). Depending on the relative timelines of SKA observations and galaxy surveys, the utility of recovered 21-cm light-cones is variable. For galaxy surveys completed prior to the availability of SKA observations, recovered light-cones may provide supplemental information to existing high-redshift galaxy catalogues. Conversely, following the operational window of SKA, recovered light-cones may be used to guide searches for galaxies located in ionized regions. In either case, characterizing the impact an imperfect U-Net recovery will have on the inferred luminosity functions is necessary.

In what follows, we will demonstrate how 21-cm images can be used in cooperation with high-redshift galaxy surveys to identify the ionization environment of galaxies during the EoR. In Section \ref{sec:2} we discuss the generation of 21-cm images, halo catalogues, and outline the introduction of instrumental and foreground effects to corrupt 21-cm images. The network architecture, training procedure, and recovery performance are presented in Section \ref{sec:3}. In Section \ref{sec:4} we discuss the halo-to-galaxy connection and explore the efficacy of recovered 21-cm light-cones in the context of high-redshift galaxy mapping. We summarize our conclusions in Section \ref{sec:5}.

\section{Simulations and Foreground Filtering}\label{sec:2}

\subsection{21-cm Brightness Temperature Fields}

To train, validate, and test our network, we generated a suite of cosmological simulations of the 21-cm brightness temperature field ($\Delta T_{21}$) over a range of redshifts during the EoR. To generate these fields, we used the Python-wrapped version, \texttt{py21cmFAST}, of the semi-numerical cosmological simulation code \texttt{21cmFASTv3} (\citealt{21cmFAST_2010, Murray_2020_JOSS}). We generated two primary simulation databases; the first consisting of 21-cm brightness temperature fields evaluated at fixed redshifts (coeval-boxes), and the second consisting of 21-cm brightness temperature fields with light-cone effects (light-cones). We use the coeval-box database as a means of control to guide our intuition regarding our network's performance on the light-cone database, and to demonstrate an improvement in performance relative to GH21. Following GH21, we fix all of our simulations to have a spatial resolution of $\Delta x = 1.5$ cMpc. We set the dimensions of our coeval-box simulation volume to be $128\times128\times128$ voxels and our light-cone simulation volume to be $128\times128\times768$ voxels. We employ \texttt{21cmFASTv3}'s default astrophysical and cosmological parameter values when generating simulations. Importantly, each simulation is generated from a different random seed, ensuring that the initial conditions of the cosmological density field are unique to each realization. It should be noted that \texttt{21cmFASTv3} uses the \cite{Parketal_v3} parameterization of galaxy properties, calibrated to high-$z$ UV luminosity functions. The details of our galaxy modeling will be described more in Section \ref{sec:4}.

Our coeval-box database consists of 780 21-cm brightness temperature boxes, distributed between $z=[4.50,10.75]$. This redshift interval corresponds to a range in the neutral hydrogen fraction, $x_{\text{HI}} \approx [0.00015,0.95]$. Of the 780 coeval boxes, 680 are distributed uniformly between $z=[6.75,8.50]$ (or $x_{\text{HI}} \approx [0.25,0.75]$) with $\Delta z = 0.25$ spacings. The remaining 100 coeval boxes are generated between $z=[4.50, 6.50]$ and $z=[8.75, 10.75]$ with $\Delta z = 0.50$ spacings, and are reserved exclusively for post-training out-of-distribution testing (see Section \ref{sec:3.2} for details). Our light-cone database consists of 350 21-cm brightness temperature light-cones, each extending from $z=5.577$ to $z=8.943$ along the LoS direction. In addition to our database of ``true'' \texttt{21cmFAST} light-cones, we also generate a secondary database of 50 approximate light-cones. Approximate light-cones are produced by concatenating coeval-boxes generated from the same initial conditions (random seed) at a series of increasing redshifts. The resulting light-cones evolve very coarsely along the LoS. The coeval-boxes comprising these light-cones have the same shape and physical dimension as those in the coeval-box database ($128\times128\times128$ voxels, $192\times192\times192$ cMpc$^3$). To produce each approximate light-cone we concatenate 6 coeval-boxes along the LoS axis, such that the final shape agrees with those in our ``true'' light-cone database ($128\times128\times768$ voxels). The constituent coeval-boxes are generated at 6 intermediate redshifts within the ``true'' light-cones, selected at 128-voxel increments along the LoS starting from $z=5.786$. This interpolation method was chosen because the resulting light-cones were found to have a $x_{\text{HI}}$ that matched closely with the $x_{\text{HI}}$ of ``true'' light-cones. The approximate light-cone database is necessary to facilitate the high-redshift galaxy recovery analysis discussed in Section \ref{sec:4} due to the intrinsic limitations of the halo-finding algorithm that is employed. We elaborate on this further in Section \ref{sec:halo_catalogues}. In Section \ref{sec:3.3}, we verify the validity of this approximation, confirming that the network's performance is robust to ``true" \textit{and} approximate light-cones.

\subsection{Halo Catalogues}\label{sec:halo_catalogues}

While the previous section addressed the generation of 21-cm images, we must also generate a corresponding database of galaxies, such that we can establish the desired connection between 21-cm imaging experiments and galaxy surveys. To facilitate this, we first consider the dark matter halos within which galaxies form (e.g. \citealt{Risa2018}). To generate dark matter halo catalogues corresponding to our 21-cm brightness temperature fields, we use the \texttt{21cmFAST} halo finder. The halo finder is run on each coeval box's density field to return a list of the masses and spatial coordinates of halos within the simulation volume. By default, the \texttt{21cmFAST} halo finder generates halo masses consistent with the Sheth-Tormen mass function (\citealt{sheth-tormen}), and assumes a turnover mass $M_{\text{Turn}}=10^{8.7} M_{\odot}$. This quantity determines the minimum mass and total number of halos identified in the simulation volume. A more in-depth explanation of the \texttt{21cmFAST} halo finder is provided in \cite{2007ApJ...669..663M}. If we consider only Population II galaxies, the reionization history is weakly sensitive to $M_{\text{Turn}}$, because the star formation efficiency inferred from UV luminosity functions falls steeply as halo mass decreases (\citealt{Jordan2017,Parketal_v3}). If, however, the physics of star formation changes in low-mass halos, and/or there are new source populations at high-$z$, then our model is subject to change (e.g. \citealt{Qin_2020, Munoz_2022, Gessey-Jones_2022}). Indeed, early JWST results hint at departures from Hubble-era model predictions, at least at $z\gtrsim 10$. 

\begin{figure*}
	\includegraphics[width=0.9\textwidth]{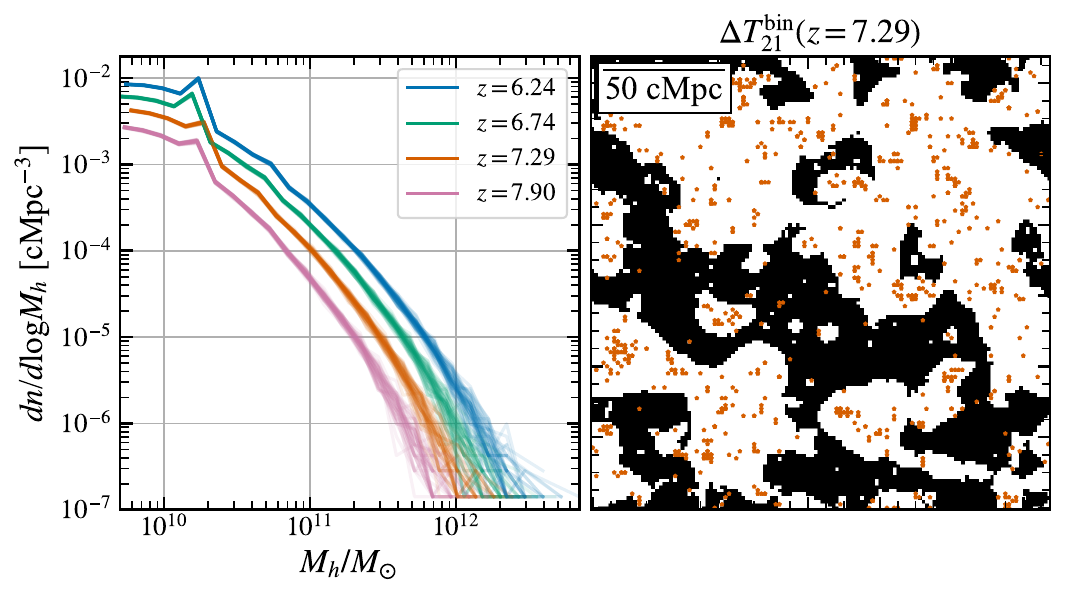}
    \caption{Left: Halo mass functions computed after running the \texttt{21cmFAST} halo finder on coeval-boxes comprising the approximate light-cones at each of the composite redshifts $z=[6.24,6.74,7.29,7.90]$. Right: A two-dimensional slice of a binarized 21-cm brightness temperature field at $z=7.29$ overlayed with the corresponding halo field (orange). When binarizing the \texttt{21cmFAST} brightness temperature field, we apply a strict binarization threshold of $0$ mK, mapping all pixels with $\Delta T_{21} > 0$ to $x_{\text{HI}}=1$. Black pixels represent regions that are completely neutral or partially ionized in the original 21-cm brightness temperature field, while white regions represent completely ionized regions in the original field. Note that nearly all halos fall in ionized regions of the map.}
    \label{fig:HMFs}
\end{figure*}

The motivation for generating the secondary approximate light-cone database discussed in the previous section is because the \texttt{21cmFAST} halo finder is configured to process coeval-boxes. Thus, we run the halo finder on an approximate light-cone's composite coeval-boxes to generate an effective light-cone halo catalogue. Figure \ref{fig:HMFs} illustrates the halo mass functions (HMFs) calculated from each of these halo catalogues, where each curve denotes a different random realization. The HMFs are grouped by the redshift of the coeval-box on which the halo finder was run. A transverse slice of a light-cone is shown in Figure \ref{fig:HMFs} overlayed with the corresponding halo catalogue. It is immediately obvious that nearly all halos fall within the ionized (white, $x_{\text{HI}}=0$) regions of the binarized \texttt{21cmFAST} brightness temperature field. This provides a preliminary indication that our methodology aimed at using ionization maps as guides for high-redshift galaxy observations is promising. The galaxy-recovery analysis we conduct in Section \ref{sec:4} investigates the specific nature of this relationship for U-Net recovered foreground wedge-removed light-cones. 

\subsection{Instrumental Noise and Foreground Removal}\label{sec:2.3}

\begin{figure}
	\includegraphics[width=\columnwidth]{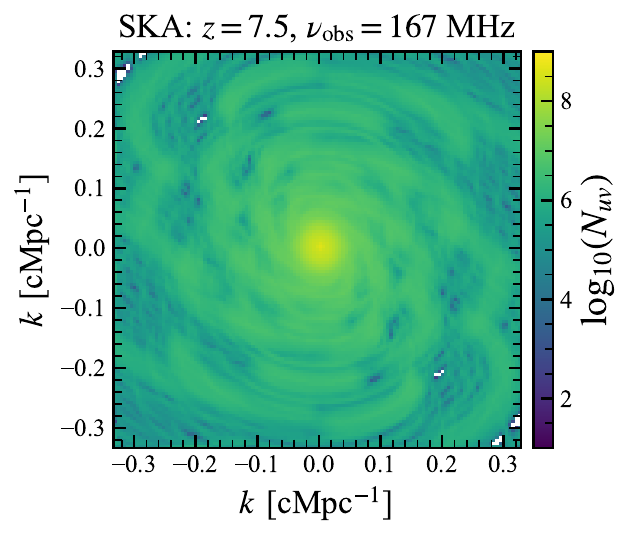}
    \caption{SKA1-Low $uv$-coverage at $167$ MHz ($z=7.5$), for $2000 \, \text{hrs}$ of total observation. The pixel intensity $N_{uv}$ denotes the number of times a particular $uv$-mode is observed. White regions indicate $uv$-modes that are not observed by the interferometer ($N_{uv}=0$). The baselines of SKA1-Low span a vast dynamic range and can thus probe a variety of $k$-modes. Given $\sigma_{uv} \propto N_{uv}^{-1/2}$, $uv$-modes with higher $N_{uv}$ will have a higher signal-to-noise. Considering the concentration of $N_{uv}$ at lower $k$ in the above plot, lower $k$-modes are less noisy than higher $k$-modes.}
    \label{fig:uv_coverages}
\end{figure}

Having generated noiseless and foreground-free coeval box and light-cone databases, we corrupt these images by introducing real-world instrumental and foreground effects. To begin, we subtract off the mean of the signal for each two-dimensional slice transverse to the LoS-axis, $\Delta \bar{T}_{21}(z)$. Doing so simulates the effect of observing the sky with an interferometer, whereby we measure fluctuations in the mean brightness temperature after discarding the null ($b=0$) baseline that corresponds to the $\mathbf{u}=0$ mode, where $\mathbf{u} \equiv (u,v)$ is the Fourier dual to angular coordinates on the sky. Adopting SKA1-Low as our fiducial instrument, we calculate its \textit{uv}-coverage and instrumental noise using the \texttt{tools21cm} Python package (\citealt{Giri_tools21cm}). The \textit{uv}-coverage is computed from the specific antenna configuration of the interferometer (SKA1-Low 2016\footnote{The latitude and longitude coordinates of the 512 antennae in the SKA1-Low configuration are presented in Appendix I of \url{https://www.skao.int/sites/default/files/documents/d18-SKA-TEL-SKO-0000422_02_SKA1_LowConfigurationCoordinates-1.pdf}}) after assuming a particular observation strategy. In this work we use an integration time of $t_{\text{int}}=10\,\text{s}$, a daily observation time of $t_{\text{obs,day}} = 6 \, \text{hrs}$, and a total observation time of $t_{\text{obs,tot}}=2000 \, \text{hrs}$. We multiply the Fourier transform of each LoS slice, $\Delta \tilde{T}_{21}(\boldsymbol{u},z)$, by the instrument's binarized $uv$-coverage (see Figure \ref{fig:uv_coverages}) at the corresponding redshift $z$, $N^{\text{bin}}_{uv}(z)$, to filter out any $uv$-modes that are not sampled by the interferometer, producing the $uv$-coverage limited 21-cm brightness temperature field $\Delta T_{21,\text{uv--lim}}(\boldsymbol{x},z)$, where $\mathbf{x}$ is the transverse position vector.

The instrumental noise for each slice along the LoS-axis is computed using a system temperature
\begin{equation}
    T_{\text{sys}}(z) = 60 \left(\frac{300 \, \text{MHz}(1+z)}{1420\, \text{MHz}}\right)^{2.55} [\text{K}].
    \label{eq:sys_temp}
\end{equation}

\noindent In $uv$-space we generate zero-mean complex Gaussian random noise with standard deviation
\begin{equation}
    \sigma_{uv}(z) = \frac{10^{26}\sqrt{2} k_B T_{\text{sys}}(z)}{A_{\text{ant}} \sqrt{\Delta \nu (z) t_{\text{int}} N_{uv}(z)}}  \, \, [\text{Jy}]
    \label{eq:uv_rms}
\end{equation}

\noindent on the sampled $uv$-modes in each LoS-slice, $\Delta \Tilde{T}_{N}(\boldsymbol{u},z)$. In Equation \eqref{eq:uv_rms}, $k_B$ is Boltzmann's constant, $A_{\text{ant}}$ is the effective antenna area, $\Delta \nu$ the frequency depth of each LoS-voxel, and $N_{uv}$ is the total number of times a given $uv$-mode is sampled over $t_{\text{obs,tot}}$ due to both redundant baselines and rotation synthesis. We then add the inverse-Fourier transform of this noise-realization, $\Delta T_{N}(\boldsymbol{x},z)$, to $\Delta T_{21,\text{uv--lim}}(\boldsymbol{x},z)$, producing the final noisy 21-cm brightness temperature field
\begin{equation}
    \Delta T_{21,\text{noisy}}(\boldsymbol{x},z) = \Delta T_{21,\text{uv--lim}}(\boldsymbol{x},z)+\Delta T_N (\boldsymbol{x},z).
    \label{eq:noisy_btemp}
\end{equation}

\noindent With the noiseless and instrument-affected coeval-boxes and light-cones on hand, we simulate the removal of foreground-contaminated Fourier modes using two different algorithms. For our coeval-box database, we follow the wedge-removal procedure outlined in GH21 and for our light-cone database, we follow the wedge-removal procedure outlined in \cite{Prelogovic_2021}. The former algorithm is as follows: 

\begin{enumerate}
 \item Fourier transform all three axes of a coeval-box,
 \item zero out Fourier modes located outside of the EoR Window (see Figure \ref{fig:foreground_wedge}),
 \item inverse Fourier transform the result.
\end{enumerate}

\noindent The latter algorithm is applied to our light-cone database (both ``true'' and approximate light-cones) and accounts for the redshift-dependence inherent to the wedge-boundary definition in Equation \eqref{eq:wedge}. This requires a recalculation of the wedge's footprint along the LoS-axis of the light-cone. Thus, for each slice along the LoS at comoving distance $r_\parallel$;

\begin{enumerate}
 \item select the section of the light-cone in the range $r_\parallel \pm \Delta r/2$ for $\Delta r = 192$ cMpc,
 \item multiply the selected light-cone by the Blackman-Harris (BH) tapering function along the LoS-axis,
 \item 3D Fourier transform the product and zero-out all Fourier modes located outside of the EoR window,
 \item inverse-Fourier transform the result, saving only the central slice.
\end{enumerate}

\noindent The foreground wedge removal algorithm outlined above reduces the dimension of light-cones along the LoS because the required $\Delta r/2$ buffer in step (i) causes an edge effect. Thus, the foreground wedge-removed light-cones that are passed to our neural network have dimension $128 \times 128 \times 512$ voxels (corresponding to a volume of $192 \times 192 \times 768 \, \text{cMpc}^{3}$). Each of the aforementioned algorithms calculate the wedge-boundary using the pessimistic assumption, $\theta_{\text{FoV}} = \pi/2$. This choice of $\theta_{\text{FoV}}$ maximizes the footprint of the wedge. Additionally, intrinsic foregrounds (see Figure \ref{fig:foreground_wedge}) are assumed to span a width 
\begin{equation}
    \Delta k_{\parallel} = \frac{2 \pi H_0 \nu_{21} E(z_{\text{min}})}{c(1+z_{\text{min}})^3(\nu_{z_{\text{min}}}-\nu_{z_{\text{max}}})}
    \label{eq:smooth_foregrounds}
\end{equation}

\noindent where $H_0$ is Hubble's constant, $\nu_{21}$ is the rest-frame emission frequency of the 21-cm line, and $\nu_{z_{\text{min}}}$ and $\nu_{z_{\text{max}}}$ are the redshifted 21-cm frequencies evaluated at the minimum and maximum redshifts $z_{\text{min}}$ and $z_{\text{max}}$ of the light-cone section considered, respectively. For the redshifts considered in this work, $ 0.05 \gtrapprox \Delta k_{\parallel} \gtrapprox 0.03$ cMpc$^{-1}$. In the case of coeval-boxes, where the redshift is constant across the entirety of the box, $\Delta k_{\parallel}$ is assumed to be $0.03$ cMpc$^{-1}$. For the particular coeval-box volumes we consider, this amounts to removing the first $k_{\parallel}$ mode.  

\begin{figure*}
    \includegraphics[width=\textwidth]{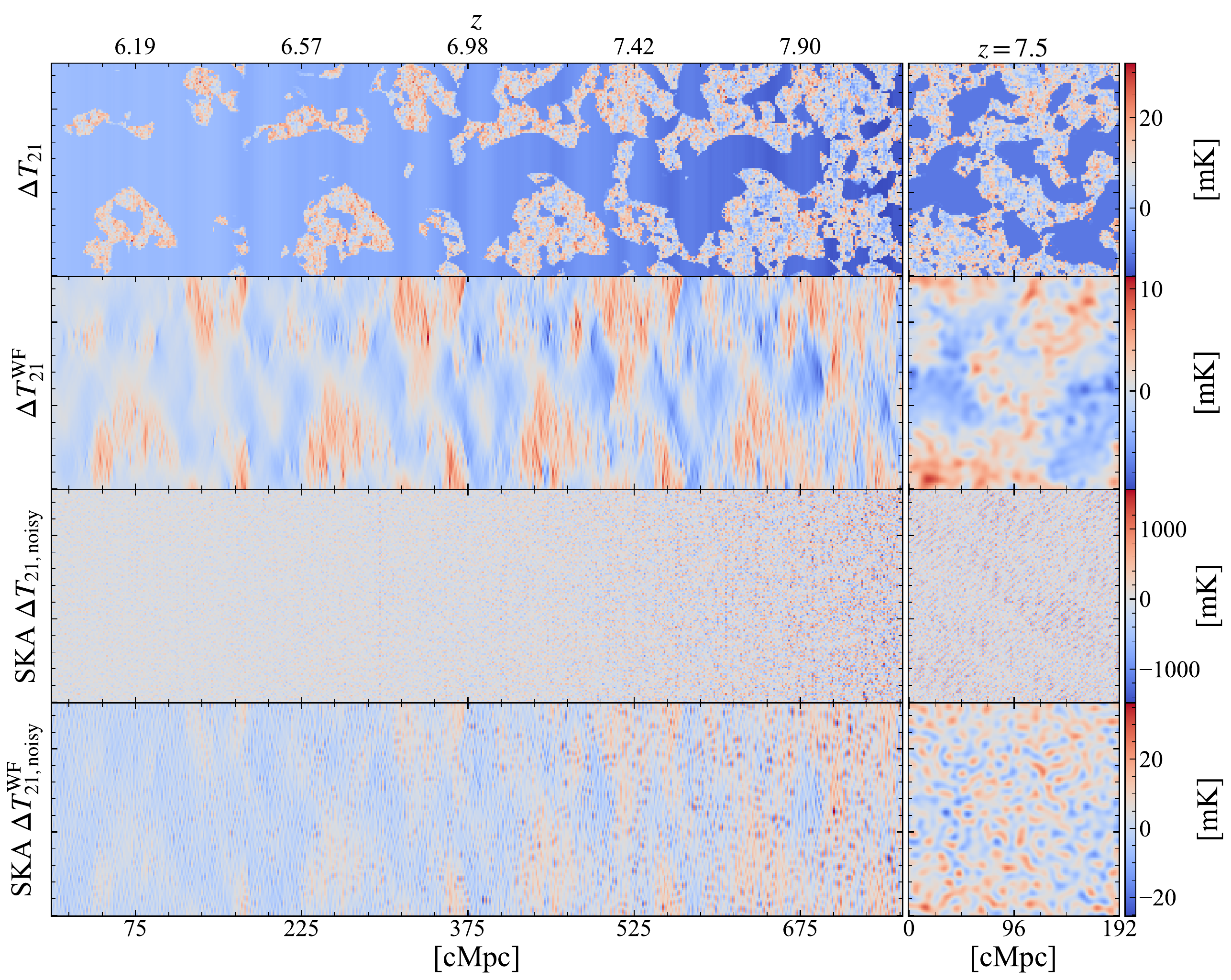}
    \caption{Sample noiseless and noisy 21-cm brightness temperature light-cones before and after foreground wedge removal. The "WF'' superscript in the second and fourth rows denote the foreground wedge-removed field. The left column displays a slice taken along the LoS-axis of the light-cone extending from $z=6.01$ to $z=8.22$. The right column displays a slice taken transverse to the LoS-axis at $z=7.5$.}
    \label{fig:wf_btemp_comparison}
\end{figure*}

Given a noisy \texttt{21cmFAST} coeval-box or light-cone, the above algorithms produce what SKA1-Low would see after foreground-contaminated wedge modes have been excised. Figure \ref{fig:wf_btemp_comparison} illustrates the distortions introduced by instrumental effects and the foreground removal procedure on \texttt{21cmFAST} light-cones. Evidently, the morphology of structures both along the LoS-axis and transverse to the LoS-axis are considerably deformed. 

\section{U-Net Recovery}\label{sec:3}

\subsection{U-Net Architecture}

\begin{figure*}
	\includegraphics[width=\textwidth]{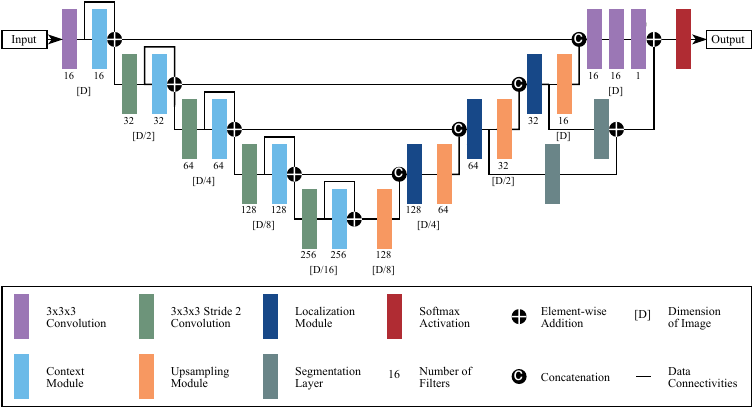}
    \caption{Block diagram of the U-Net. The dimension of the input image D is reduced in the downwards (left) path and increased in the upwards (right) path.}
    \label{fig:U-Net}
\end{figure*}

The neural network architecture employed in this work is the same as the U-Net presented in GH21, which in turn draws heavily from the architecture presented in \cite{isensee}. A schematic of the U-Net we use is shown in Figure \ref{fig:U-Net}. The context modules (shown in light blue) in the downwards (left) path of the U-Net consist of successive $3\times3\times3$ convolutional layers. In the upwards (right) path of the U-Net, upsampling modules (shown in orange) consist of a three-dimensional upsampling layer followed by a $3\times3\times3$ convolutional layer, localization modules (shown in dark blue) consist of a $3\times3\times3$ convolutional layer followed by a $1\times1\times1$ convolutional layer, and segmentation layers (shown in grey) consist of a three-dimensional upsampling layer.

As a reminder, the problem our U-Net is configured to solve is one of two-class image segmentation. A well-trained network should thus be able to identify each voxel within a foreground wedge removed 21-cm map as neutral or ionized (1 or 0). While our network is not designed to produce exact binary outputs, a binarization filter is used when evaluating predictions on the testing set. Such external binarization was implemented to incentivize the network to produce near-binary outputs. We emphasize that this binarization filter is not part of the U-Net architecture, and is therefore not present in Figure \ref{fig:U-Net}. As in GH21, we find that recovery performance is insensitive to the binarization threshold (changing minimally when varying the threshold from 0 to 1), confirming that our network has been successfully incentivized. We implement the same binarization threshold of 0.9 as GH21 on the basis that it classifies even slightly ionized regions (as determined by our U-Net) as completely ionized. This process maps all voxels in the prediction with a value greater than or equal to 0.9 to 1 and less than 0.9 to 0.

During training, we pass normalized (between 0 and 1) foreground wedge removed 21-cm maps as inputs and binarized 21-cm maps as labels to our network. For this binarization of labels (i.e., separate from the binarization of the predictions discussed previously), we use a threshold of $\Delta T_{21} = 0\, \text{mK}$, such that neutral voxels, where  $\Delta T_{21} > 0$, have a value of 1 and ionized voxels, where $\Delta T_{21} = 0$, have a value of 0. These binarized 21-cm maps act as the ground-truth comparison for the network, and are used to compute the loss during training and the recovery prediction statistics after training. We train our network using a modified binary dice coefficient (BDC) loss function
\begin{equation}\label{eq:loss}
\mathcal{L}_{\text{BDC}}(G,P) = - \frac{2|G \cap P|+\varepsilon}{|G|+|P|+\varepsilon},
\end{equation}

\noindent where $G$ and $P$ are the ground-truth and prediction arrays, $\varepsilon$ is an additive parameter for numerical stability, and $|\dots|$ denotes the cardinality of the respective array. During training $\varepsilon$ is set to $1$, which is comparatively small relative to the size of the arrays we are working with (e.g. $|G| \approx 10^7$). The BDC measures the voxel-wise agreement between the ground-truth and prediction and has demonstrated considerable utility in the fields of computer vision and biomedical image segmentation (\citealt{milletari2016vnet, Jadon_2020}). 

As a departure from the network configuration presented in GH21, we introduce minor modifications after performing hyperparameter tuning on a few key network parameters. These included the rate of spatial dropout in context modules, the degree of $L_2$ regularization in convolutional layers, and the batch size. While the effects of spatial dropout were also explored in GH21, we introduced $L_2$ regularization to combat model overfitting and improve training stability. This regularization strategy penalizes larger model weights $\boldsymbol{\omega} = (\omega_1,\omega_2,\dots,\omega_i)$, driving them closer to the origin by adding a regularization term $\Omega = \alpha_{L_2} ||\boldsymbol{\omega}||^2_2$ to the loss function (\citealt{GoodBengCour16}), where $\alpha_{L_2}$ is the $L_2$ coefficient. Accounting for this strictly positive additive regularization term, the loss computed during training (and reported in Figure \ref{fig:loss_curves}) is $\mathcal{L} = \mathcal{L}_{\text{BDC}}+\Omega$, and is therefore no longer bounded above by 0. At early epochs, a large $\Omega$ can thus overwhelm $\mathcal{L}_{\text{BDC}}$, producing a large positive $\mathcal{L}$. We found that applying $L_2$ regularization with $\alpha_{L_2} = 0.01$ only to the kernel (not the bias) of our U-Net's convolutional layers, implementing a spatial dropout rate of 0.3, and a batch size of 3, produced the best results, minimizing the loss function and stabilizing the learning process.

\subsection{U-Net Training}\label{sec:3.2}

\begin{figure}
	\includegraphics[width=\columnwidth]{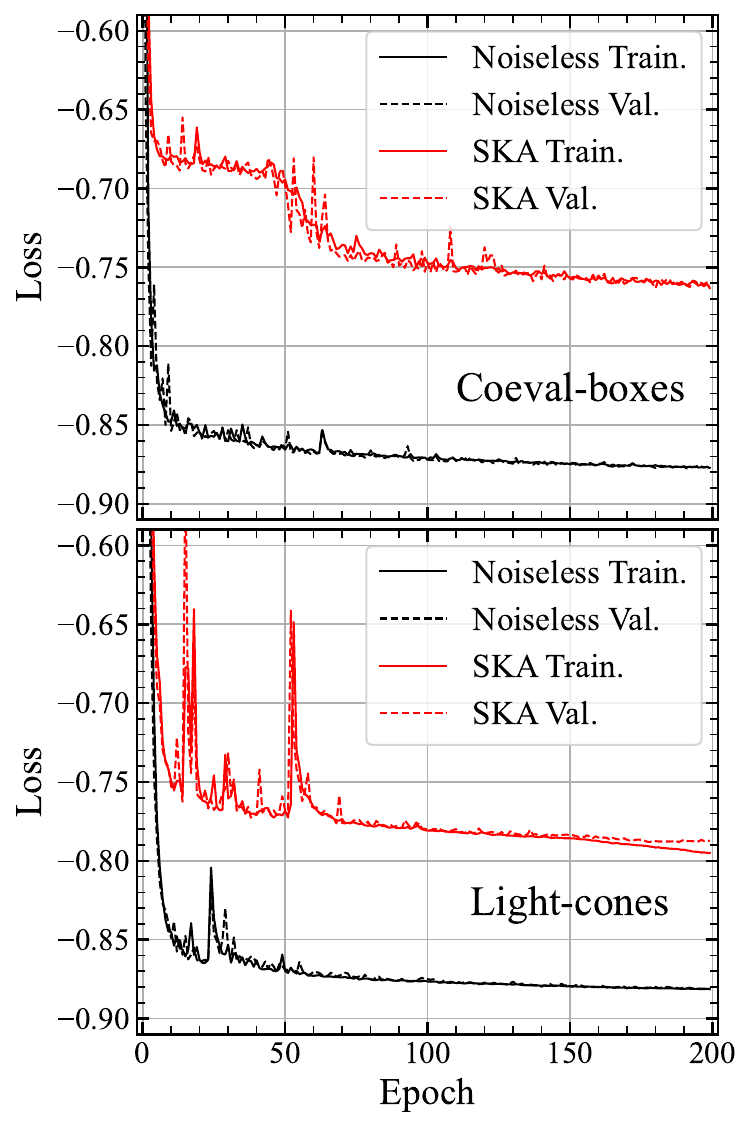}
    \caption{Upper: Training (solid) and validation (dashed) loss curves for each of the coeval-box datasets. Lower: Training (solid) and validation (dashed) loss curves for each of the light-cone datasets. The lack of a significant offset between any pair of training and validation curves is indicative of a well-trained, generalized model. The lack of any appreciable gradient in the slope of the loss curves at later epochs suggests we have hit the limitations of our network architecture and training regimen. Note that we have implemented an upper limit on the $y$-axes that masks the vertical extension of the learning curves where $\mathcal{L}>0$. This was done to reduce the dynamic range of the plot, increasing the scale of the relevant behaviour visible at later epochs.}
    \label{fig:loss_curves}
\end{figure}

For each of the different datasets discussed in Section \ref{sec:2}; (i) noiseless coeval-boxes, (ii) SKA-noisy coeval-boxes, (iii) noiseless ``true" light-cones, and (iv) SKA-noisy ``true" light-cones, we run a separate training. We employ a training to validation to test set split of 520:80:180 for coeval box datasets, and 250:50:50 for light-cone datasets. This choice of training set size was limited by the available computational resources, and thus improved network performance may be realized with considerably larger training sets (e.g. \citealt{8599448}). The test sets are not shown to the network during training and are used to provide an unbiased evaluation of the network's performance \textit{after} training. Similarly, the approximate light-cones belonging to each of the noiseless and noisy data suites are intended purely for post-training testing, and thus we do not run separate trainings on these datasets. We emphasize that the coeval box test sets include 100 coeval boxes uniformly distributed across 10 redshifts that are \textit{not} included in the training and validation sets. These boxes span the early and late stages of reionization, and are intended to evaluate the network's ability to generalize to out-of-distribution samples. The trainings for datasets (i)-(iv) are scheduled for $200$ epochs and conducted using an Adam optimizer with an exponentially decaying learning rate schedule, $lr(E) = 0.0005(0.985^E)$ for $E$ the number of elapsed epochs. The BDC loss-curves evaluated during training are shown in Figure \ref{fig:loss_curves}. The gradients of each training and validation loss curve pair in Figure \ref{fig:loss_curves} decline considerably towards the final epochs, indicating the model has become sufficiently trained. While the slight separation between the training and validation loss curves of the SKA-noisy light-cone dataset is indicative of minor overfitting (for $E>160$), our choice of 200 epochs prevents this offset from expanding further.

\subsection{U-Net Predictions}\label{sec:3.3}

\begin{figure*}
	\includegraphics[width=\textwidth]{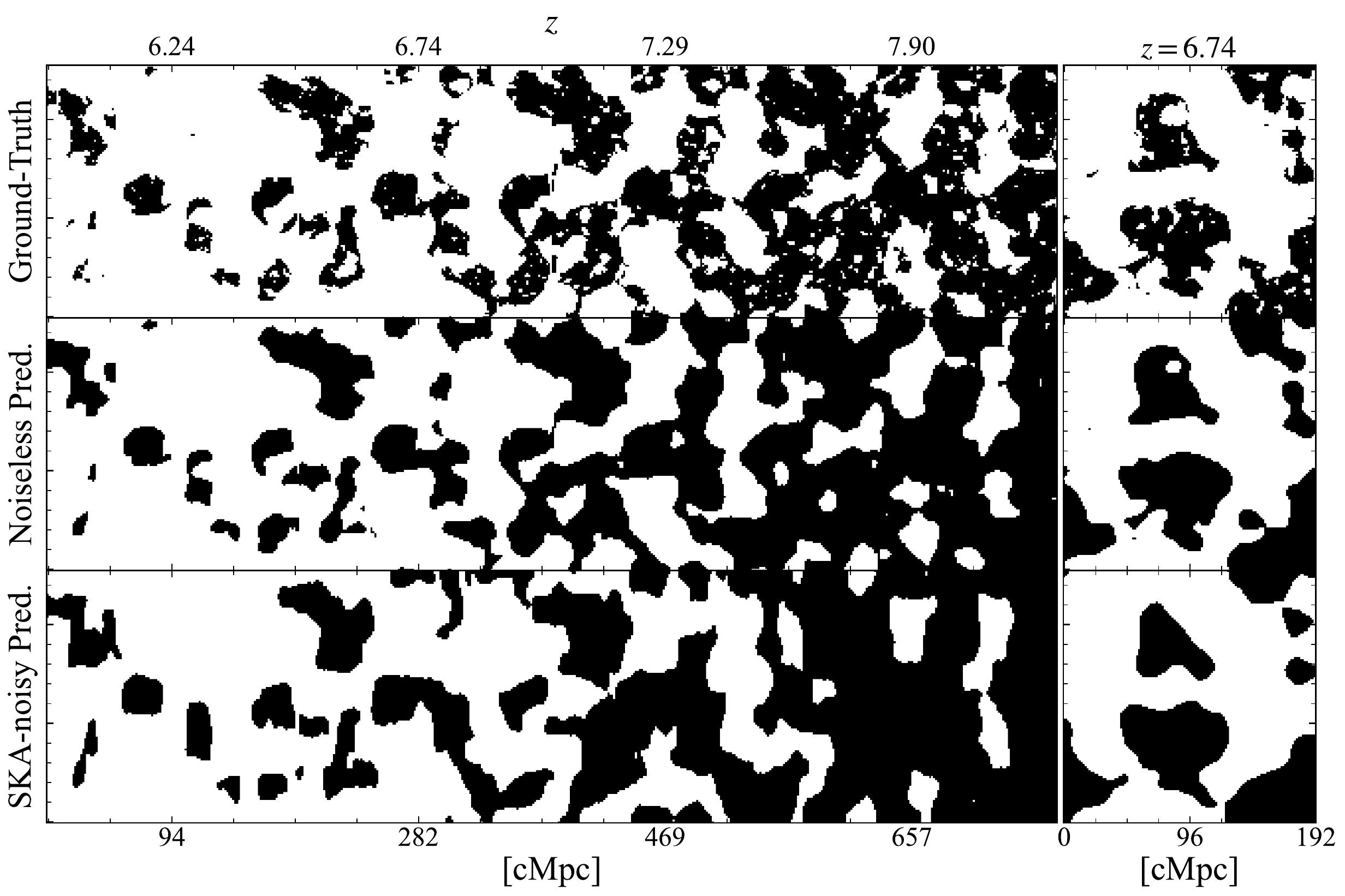}
    \caption{Example network predictions along the LoS-axis (first column) and transverse to the LoS-axis (second column) for noiseless and SKA-noisy approximate light-cone datasets. The second and third rows show the U-Net's binarized predictions for the foreground-wedge removed noiseless and SKA-noisy light-cones, respectively. It is clear that our network has successfully reconstructed the general form of the large-scale structure in the ground-truth light-cone. Notably, the network is not able to reconstruct the correct shape and location of smaller scale ionized bubbles.}
    \label{fig:fake_lc_preds}
\end{figure*}

Sample two-dimensional slices of the U-Net's three-dimensional predictions on the noiseless and SKA-noisy approximate light-cone test sets are displayed in Figure \ref{fig:fake_lc_preds}. One should note that the noiseless \textit{and} noisy datasets have the same noiseless ground-truth binarized 21-cm brightness temperature maps during training. Thus, all comparisons during training are made in reference to the same ground-truth, independent of the noise characteristics of the particular dataset. 

To quantify the U-Net's recovery performance on each test set we use accuracy, precision, recall, and mean intersection-over-union (mean IoU) as metrics. To compute these statistics, we apply the binary classification scheme in Table \ref{tab:voxel_classification} to the voxels of our predicted maps. 

\begin{table}
\centering
 \begin{tabular}{lcc}
  \hline
  Ground truth & Prediction & Classification\\
  \hline
  Ionized & Ionized & True Positive (TP) \\
  Ionized & Neutral & False Negative (FN) \\
  Neutral & Ionized & False Positive (FP) \\
  Neutral & Neutral & True Negative (TN) \\
  \hline
 \end{tabular}
 \caption{Voxel classification scheme used to compute performance statistics. Ionized or neutral refers to the binarized state of the voxel in the ground-truth or prediction.}
 \label{tab:voxel_classification}
\end{table}

\begin{figure*}
	\includegraphics[width=\textwidth]{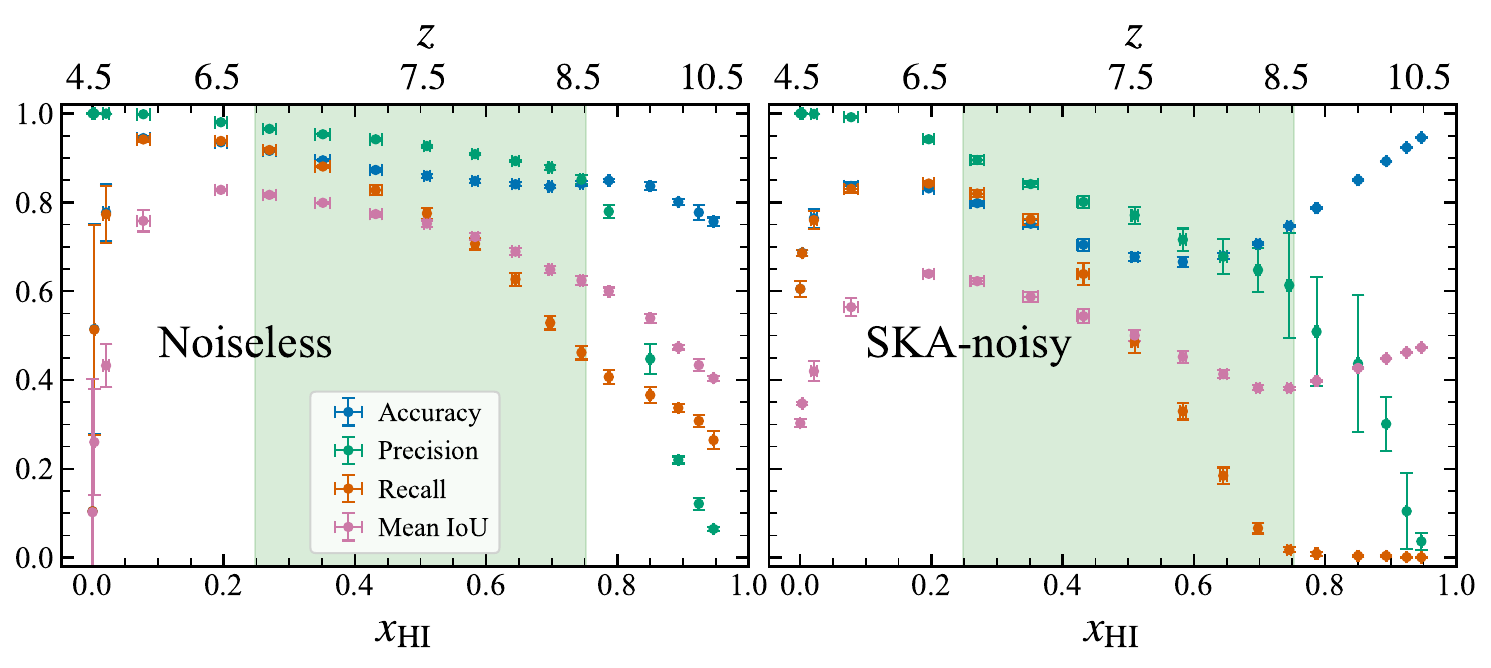}
    % paper_coeval_SKA_vs_noiseless_test_set_recovery_stats_new_noise_new_color_palette_updated_patience_100_
    \caption{The accuracy, precision, recall, and mean intersection-over-union (mean IoU) computed using Equations \eqref{eq:accuracy}--\eqref{eq:IoU} for the noiseless and SKA-noisy coeval-box test sets as a function of neutral fraction (and redshift). The green shaded region denotes the neutral fraction interval on which the network was trained, while data points outside of this region represent out-of-distribution samples used to evaluate the generalizability of the U-Net. Each data point represents the mean over 10 coeval-box realizations at the same redshift. Each data point is plotted with $1\sigma$ error bars. There is an evident drop-off in recovery performance as a function of increasing neutral fraction for both the noiseless and SKA-noisy test-sets.}
    \label{fig:coeval_rec_stats}
\end{figure*}

\begin{figure*}
	\includegraphics[width=\textwidth]{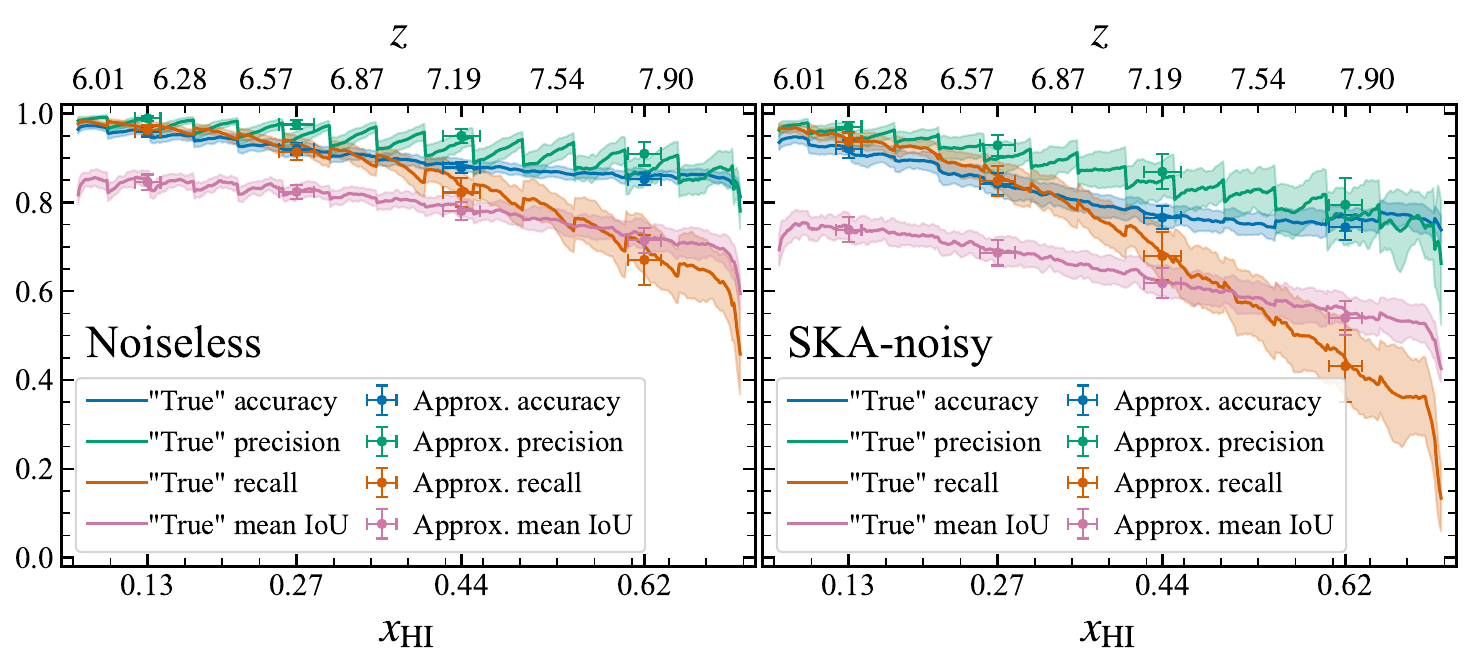}
    \caption{The accuracy, precision, recall, and mean intersection-over-union (mean IoU) computed using Equations \eqref{eq:accuracy}--\eqref{eq:IoU} for the noiseless and SKA-noisy light-cone ``true" (solid curves) and approximate (circular markers) test sets as a function of neutral fraction. The shaded regions indicate the $1\sigma$ error bands for each ``true" statistic (computed over 50 light-cones). The sawtooth structure of the curves arises due to the effects of binarizing light-cone slices that are generated by interpolating between successive coeval-boxes in \texttt{21cmFAST}.}
    \label{fig:lightcone_rec_stats}
\end{figure*}

\begin{figure*}
	\includegraphics[width=0.8\textwidth]{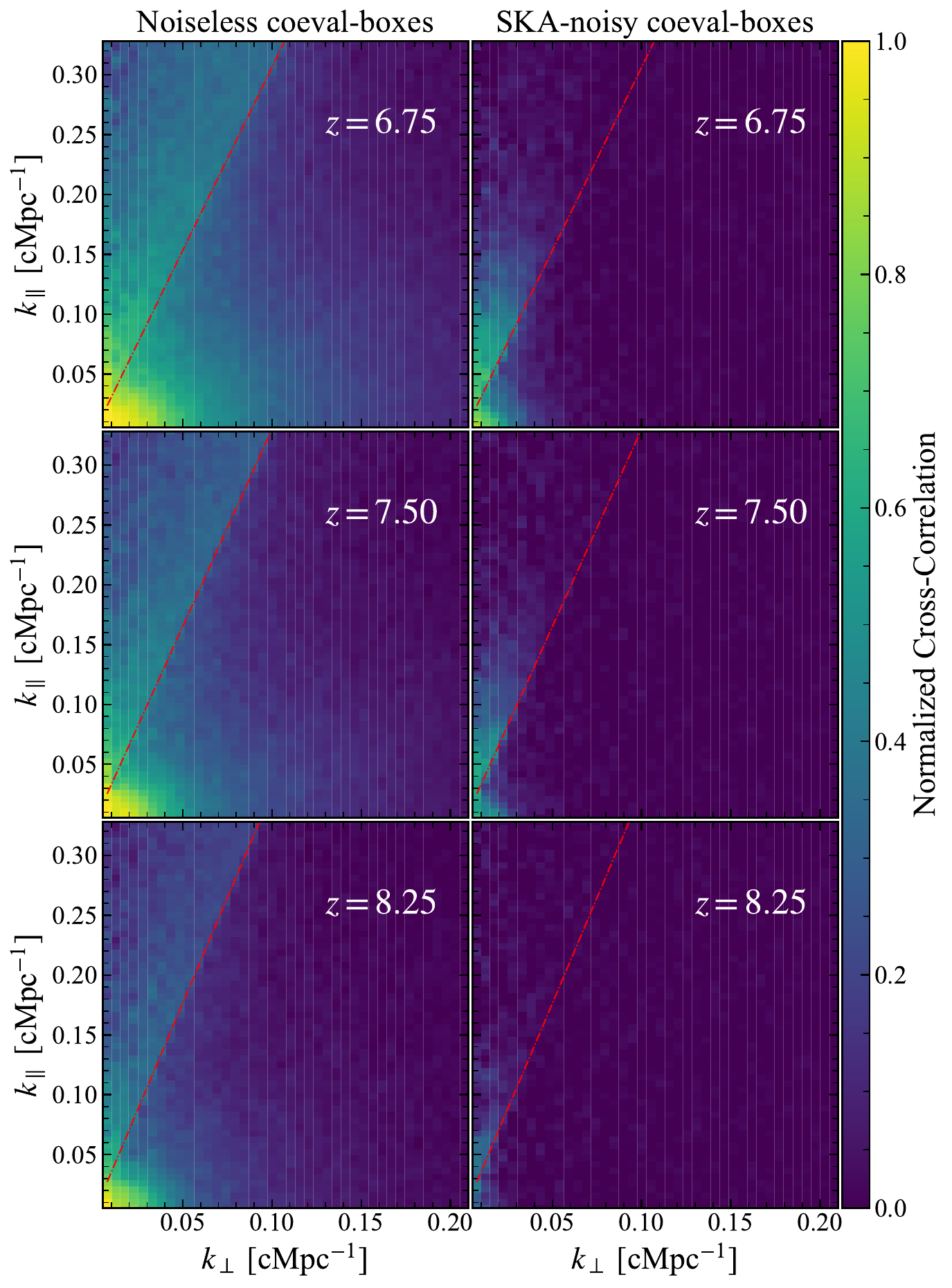}
    \caption{Mean normalized cross-power of the binarized ground-truth and U-Net predicted coeval-boxes at redshifts $6.75, 7.50, $ and $8.25$. The boundary of the foreground-wedge computed at each redshift using Equation \eqref{eq:wedge} is overlayed in red. Note that we have limited the extent of the $k_{\perp}$-axis to $\sim 0.2 \,\text{cMpc}^{-1}$ given $\mathcal{N} \sim 0$ for all modes beyond this cutoff.}
    \label{fig:norm_CrossCorr_coeval}
\end{figure*}

\begin{figure*}
	\includegraphics[width=0.8\textwidth]{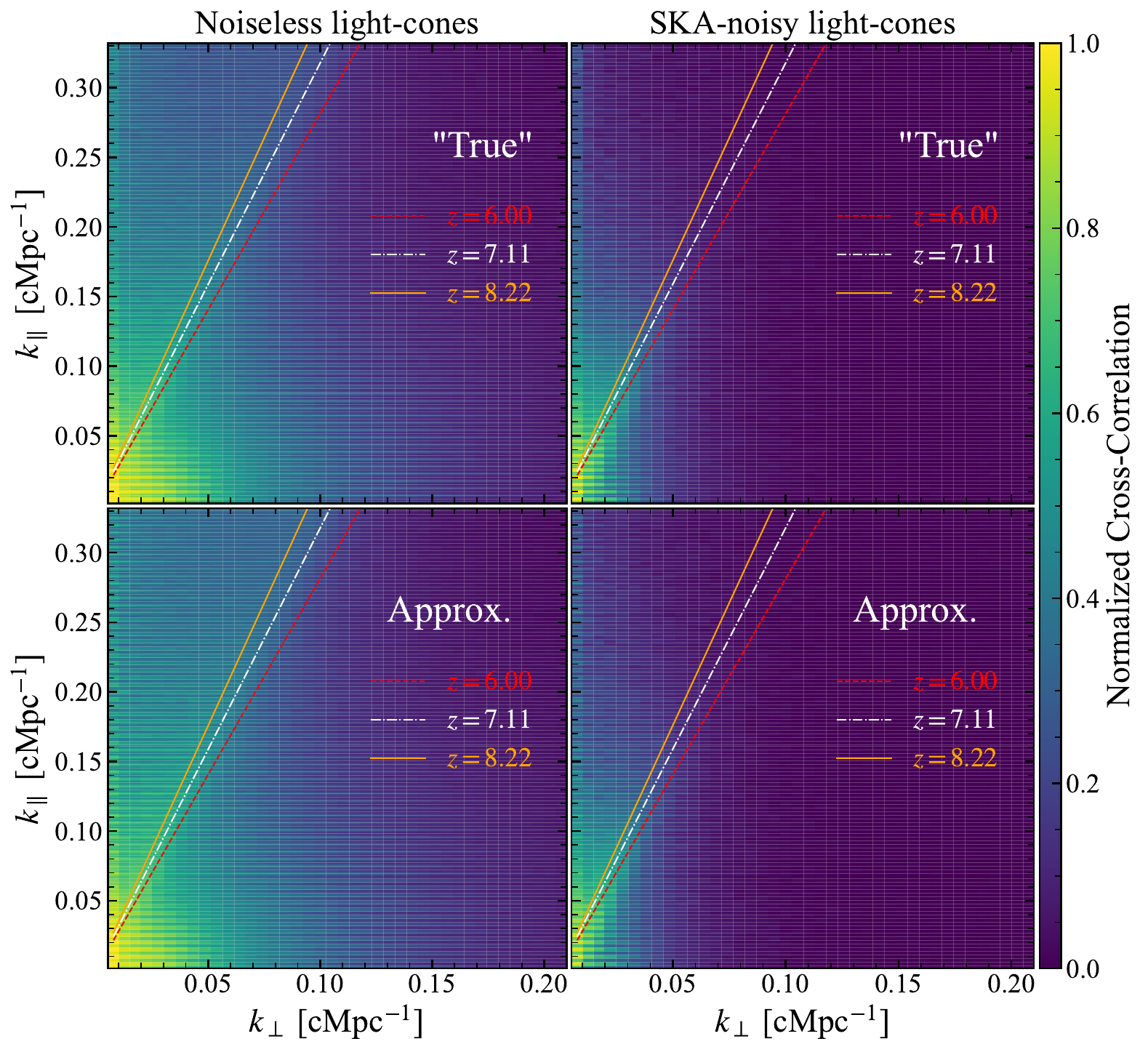}
    \caption{Mean normalized cross-power of the binarized ground-truth and U-Net predicted light-cones for both the ``true" and approximate light-cone test-sets. The boundary of the foreground-wedge for redshifts $6.00, 7.11,$ and $8.22$ are overlayed in red, white, and orange, respectively. These redshifts represent the minimum, median, and maximum wedge-angle used to excise the foreground wedge from light-cones using the algorithm described Section \ref{sec:2.3}. As in Figure \ref{fig:norm_CrossCorr_coeval}, we have again limited the extent of the $k_{\perp}$-axis to $\sim 0.2 \, \text{cMpc}^{-1}$ given $\mathcal{N} \sim 0$ for all modes beyond this cutoff.}
    \label{fig:norm_CrossCorr_lightcones}
\end{figure*}

Using the classification presented in Table \ref{tab:voxel_classification}, the accuracy is defined as 
\begin{equation}
\text{Accuracy} = \frac{\text{TP}+\text{TN}}{\text{TP}+\text{FP}+\text{FN}+\text{TN}} = \frac{\text{$N_{\text{vox}}$ correctly labelled}}{\text{$N_{\text{vox}}$}}
\label{eq:accuracy}
\end{equation}
and indicates how often the predictions match their labels, where $N_{\text{vox}}$ is the number of voxels. Precision is defined as 
\begin{equation}
\text{Precision} = \frac{\text{TP}}{\text{TP}+\text{FP}} = \frac{\text{$N_{\text{vox}}$ correctly labelled as ionized}}{\text{$N_{\text{vox}}$ ionized in prediction}}
\label{eq:precision}
\end{equation}
and measures how many voxels the network labels as ionized are ionized in the ground-truth. %Equivalently, one may consider precision to be a measure of how frequently the network misidentifies a neutral region as being ionized. 
Recall is a measure of the number of ground-truth ionized voxels that are labelled as ionized by the network 
\begin{equation}
\text{Recall} = \frac{\text{TP}}{\text{TP}+\text{FN}} = \frac{\text{$N_{\text{vox}}$ correctly labelled as ionized}}{\text{$N_{\text{vox}}$ ionized in ground truth}}.
\label{eq:recall}
\end{equation}
The final statistic, mean IoU, quantifies the degree of overlap between the ground-truth and prediction
\begin{equation}
\text{mean IoU} = \frac{1}{2}\left(\frac{\text{TP}}{\text{TP}+\text{FN}+\text{FP}}+\frac{\text{TN}}{\text{TN}+\text{FN}+\text{FP}}\right).
\label{eq:IoU}
\end{equation}
We compute these statistics over each of the four test sets described in Section \ref{sec:3.2} as well as the two approximate light-cone datasets (noiseless and noisy). Figures \ref{fig:coeval_rec_stats}  and \ref{fig:lightcone_rec_stats} tabulate the value of each statistic as a function of neutral fraction and redshift. Given the steadily increasing accuracy, precision, recall, and mean IoU curves in Figures \ref{fig:coeval_rec_stats} and \ref{fig:lightcone_rec_stats}, it is evident that the overall recovery performance of our network increases as redshift decreases, a trend consistent with that reported in \cite{Hassan_new_citation}. In Figure \ref{fig:coeval_rec_stats}, we see reasonable network performance beyond the training and validation set redshifts (outside of the green shaded regions), although this deteriorates towards the extremes where the reconstruction of tiny ionized bubbles at the beginning of reionization is very difficult (and similarly for tiny neutral islands when reionization is almost entirely complete). The redshift-dependent performance between these extremes can be explained intuitively by considering how the structures of interest in binarized 21-cm maps evolve with redshift. As reionization progresses, smaller, isolated bubbles merge to form larger ionized bubbles. These larger structures are more easily learned by our network, and thus are more consistently reproduced in predicted maps.
Therefore, our network is well-suited to identify and reconstruct the location and general morphology of the largest ionized bubbles in the simulation volumes. This conclusion is quantitatively supported by considering the normalized cross-power spectra of the ground-truth and prediction arrays, defined as
\begin{equation}
    \mathcal{N}(k_{\perp},k_{\parallel}) = \frac{\mathcal{G}\mathcal{P}^*}{\sqrt{(\mathcal{G}\mathcal{G}^*)(\mathcal{P}\mathcal{P}^*)}}
    \label{eq:norm_crosscorr}
\end{equation}
where $\mathcal{G}$ and $\mathcal{P}$ are the three-dimensional Fourier transforms of the ground-truth and prediction arrays $G$ and $P$, respectively. Implicit in this equation is the binning of each product (whether $\mathcal{G} \mathcal{P}^*$, $\mathcal{G} \mathcal{G}^*$, or $\mathcal{P} \mathcal{P}^*$) into $(k_{\perp},k_{\parallel})$ bins. In the case of a perfect reconstruction, where $G=P$, $\mathcal{N}=1$ for all $(k_{\perp},k_{\parallel})$. The two-dimensional normalized cross-power spectrum provides a means of quantifying the fidelity of the network's recovery on different spatial scales along different axes. Further, because we excise all modes lying within the foreground wedge, $\mathcal{N}(k_{\perp},k_{\parallel})$ demonstrates explicitly how well our network is able to recover excised modes. Figures \ref{fig:norm_CrossCorr_coeval} and \ref{fig:norm_CrossCorr_lightcones} show the normalized cross-power spectra for the noiseless and SKA-noisy coeval-box and light-cone predictions, respectively. The boundary of the foreground wedge is plotted overtop the spectra, such that all $(k_{\perp},k_{\parallel})$-modes lying between the boundary and the $k_{\perp}$-axis are reconstructed by the U-Net. There is clearly a successful reconstruction at some level (especially at low $k \equiv \sqrt{k_\perp^2 + k_\parallel^2}$). While the recovered modes certainly are not perfect (and at very high $k$ there is essentially no recovery at all), Figure~\ref{fig:fake_lc_preds} shows that the limited recovery is at least enough to accurately reconstruct the larger ionized regions.

The accentuated drop-off in recovery performance towards higher redshifts for the noisy datasets in Figures \ref{fig:coeval_rec_stats} and \ref{fig:lightcone_rec_stats}, is also noticeable in Figure \ref{fig:norm_CrossCorr_coeval}, and is due in part to the compounding effect of a declining signal to noise ratio (SNR). This is a result of the redshift dependence inherent to Equation \eqref{eq:uv_rms}, and further explains the expanding offset between the performance statistics of the noiseless and noisy test sets in Figures 
\ref{fig:coeval_rec_stats} and \ref{fig:lightcone_rec_stats}, and the larger discrepancies between the noiseless and noisy spectra in Figure \ref{fig:norm_CrossCorr_coeval} at higher redshifts. 

One should note that the prominent sawtooth structure apparent in Figure \ref{fig:lightcone_rec_stats} arises due to the effects of binarizing light-cone slices that are generated by interpolating between successive coeval-boxes within the \texttt{21cmFAST} light-cone generation algorithm. This structure is therefore not an artefact of the U-Net recovery, but rather a product of the binarization filter necessary to compute the recovery statistics.

One important takeaway from Figures \ref{fig:lightcone_rec_stats} and \ref{fig:norm_CrossCorr_lightcones} is that the network's recovery performance over the approximate light-cone datasets is consistent with the "true" light-cone datasets. This indicates that we have reasonably captured some of the more salient characteristics of the standard \texttt{21cmFAST} light-cone when constructing light-cones from coeval-boxes with very coarse redshift spacings.

In summary, we have demonstrated that our U-Net can successfully reconstruct foreground wedge-removed coeval boxes across a wide range of redshifts. As a notable advancement to GH21, we have extended the U-Net's capability to process light-cones, which incorporate the redshift evolution that will be implicit in real observations. Further, we have demonstrated that the U-Net's recovery is still reliable when instrumental limitations and noise are accounted for.

\section{Galaxy Recovery}\label{sec:4}

Using the recovered approximate light-cones along with their corresponding halo catalogues, we now shift to a discussion of galaxy recovery. As alluded to previously, the utility of recovered light-cones is two-fold; (1) for galaxy surveys completed prior to the availability of tomographic 21-cm datasets, recovered light-cones will supplement existing galaxy catalogues with information regarding the ionization state of the surveyed regions, (2) once tomographic 21-cm datasets are available, light-cones may guide searches for galaxies in ionized regions. For example, if a collection of ionized regions are identified where galaxies have yet to be detected, proposals for follow-up observations with higher sensitivity may be scheduled to probe for lower luminosity galaxies. The size of recovered ionized regions may also be used to prioritize spectroscopic follow-up. For example, given Lyman-$\alpha$ emitters are expected to reside in large ionized regions during the EoR (e.g. \citealt{Furlanetto_2004_LAEs}), one may prioritize these regions for follow-up given their importance as a highly sensitive probe of reionization (e.g. \citealt{Haiman_2002,McQuinn_2007_LAEs}). 

With respect to (1), it is of interest to quantify how well the galaxy luminosity function (LF) measured in the predicted ionized regions matches the LF of galaxies located in the ground-truth ionized regions. To summarize the implications an imperfect light-cone recovery will induce on the inferred galaxy LF, we compute a multiplicative correction factor to map from the inferred to the true galaxy LF. 

\begin{figure}
	\includegraphics[width=\columnwidth]{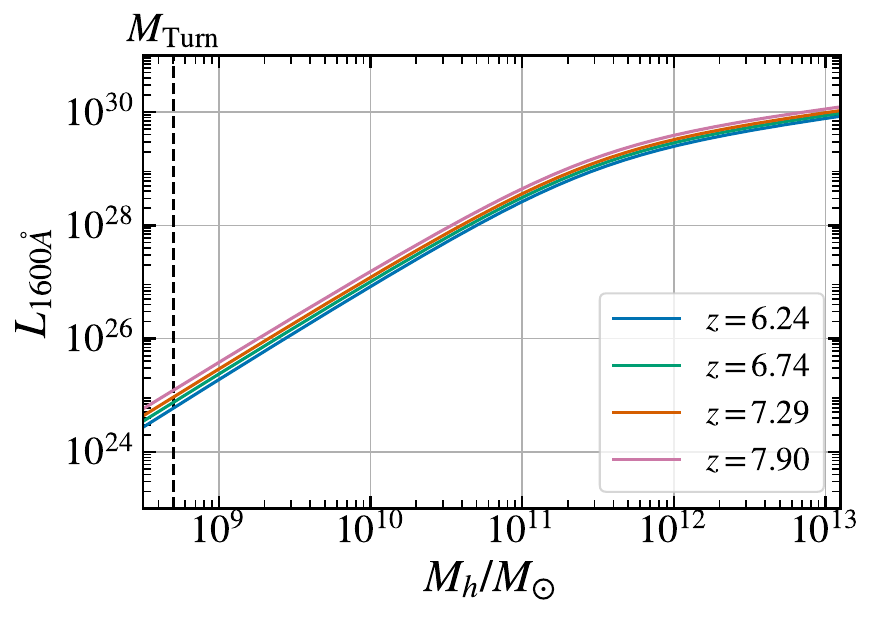}
    \caption{UV luminosity to halo mass relation used to convert halo catalogues into galaxy catalogues. The dashed vertical line denotes the turnover mass used by the \texttt{21cmFAST} halo finder ($10^{8.7} M_{\odot}$).}
    \label{fig:lumo_mass_curves}
\end{figure}

To perform this analysis, we adopt the relationship between halo mass and rest-UV luminosity in \cite{Jordan2017} (see Figure \ref{fig:lumo_mass_curves}), a semi-empirical model calibrated to high-z luminosity functions from \cite{Bouwens2015}, consistent with the \cite{Parketal_v3} parameterization employed in \texttt{21cmFASTv3}. The model assumes a double power-law relationship between halo mass and the efficiency of star formation, and a constant conversion factor between star formation rate and UV luminosity from the \texttt{BPASS} version 1.0 models (\citealt{Eldridge2009}) for solar metallicity stars forming at a constant rate for $100\,\text{Myr}$. Dust reddening is neglected in these models for simplicity.

Having converted the halo catalogues into galaxy catalogues using the aforementioned relation, we sort galaxies using an analogous classification scheme to that presented in Table \ref{tab:voxel_classification}:

\begin{enumerate}
 \item true positive galaxies (TP$_{\text{gal}}$) are located in ionized voxels of the ground-truth and ionized voxels of the prediction,
 \item false negative galaxies (FN$_{\text{gal}}$) are located in ionized voxels of the ground-truth and neutral voxels of the prediction,
 \item false positive galaxies (FP$_{\text{gal}}$) are located in neutral voxels of the ground-truth and ionized voxels of the prediction,
 \item true negative galaxies (TN$_{\text{gal}}$) are located in neutral voxels of the ground-truth and neutral voxels of the prediction.
\end{enumerate}

\noindent Using this classification scheme, we define the following LFs:

\begin{enumerate}
 \item the ground truth ionized (GTI) LF, $\Phi_{\text{GTI}}$, as the LF of galaxies located in ionized voxels of the ground-truth (TP$_{\text{gal}}$+FN$_{\text{gal}}$),
 \item the ground truth neutral (GTN) LF, $\Phi_{\text{GTN}}$, as the LF of galaxies located in neutral voxels of the ground-truth (TN$_{\text{gal}}$+FP$_{\text{gal}}$),
 \item the predicted ionized (PI) LF, $\Phi_{\text{PI}}$, as the LF of galaxies located in ionized voxels of the prediction (TP$_{\text{gal}}$+FP$_{\text{gal}}$),
 \item the predicted neutral (PN) LF, $\Phi_{\text{PN}}$, as the LF of galaxies located in neutral voxels of the ground truth (TN$_{\text{gal}}$+FN$_{\text{gal}}$),
 \item the global LF, $\Phi_{\text{Global}}$, to be the LF of all galaxies (TP$_{\text{gal}}$+TN$_{\text{gal}}$+FP$_{\text{gal}}$+FN$_{\text{gal}}$), irrespective of their location in an ionized or neutral voxel.
\end{enumerate}

\begin{figure}
	\includegraphics[width=\columnwidth]{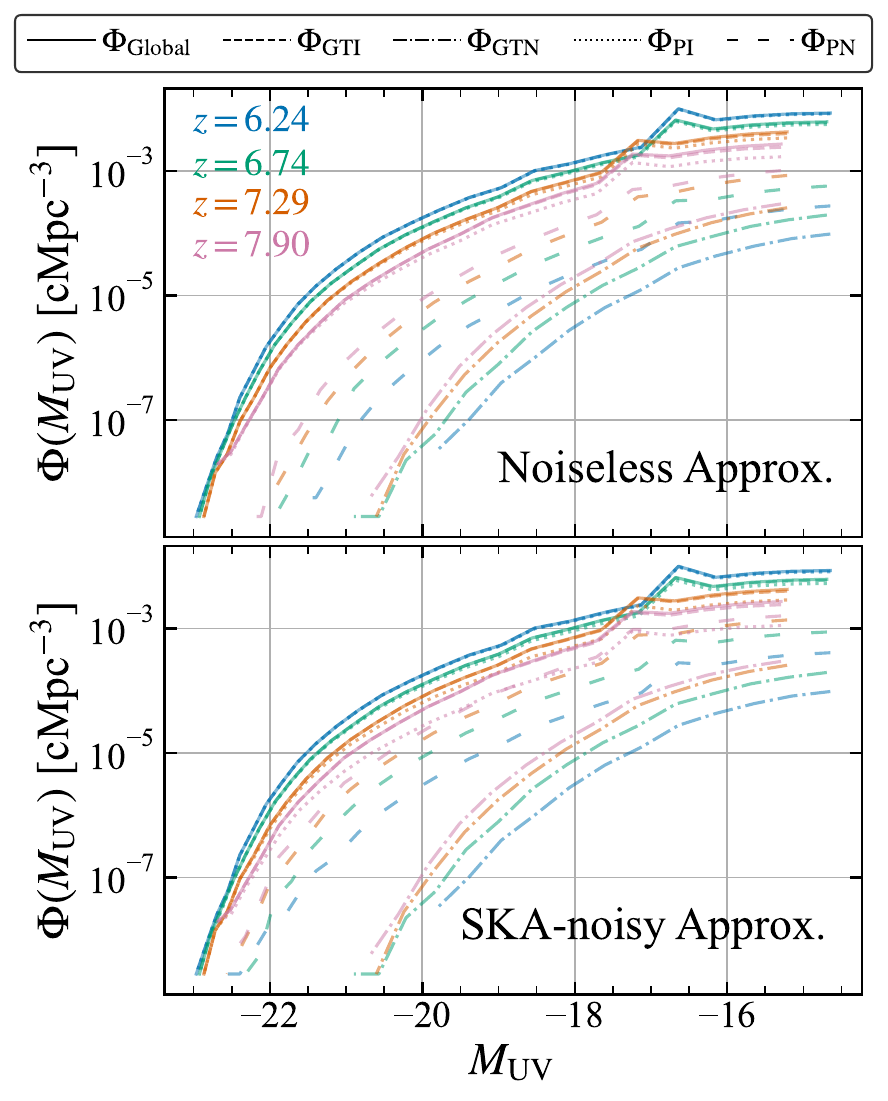}
    \caption{Global (solid line), GTI (dashed line), GTN (dash-dotted line), PI (dotted line), and PN (loosely-dashed line) LFs for the noiseless and SKA-noisy approximate light-cones at redshifts $6.24, 6.74, 7.29$ and $7.90$. }
    \label{fig:LFs_curves}
\end{figure}

\begin{figure}
	\includegraphics[width=\columnwidth]{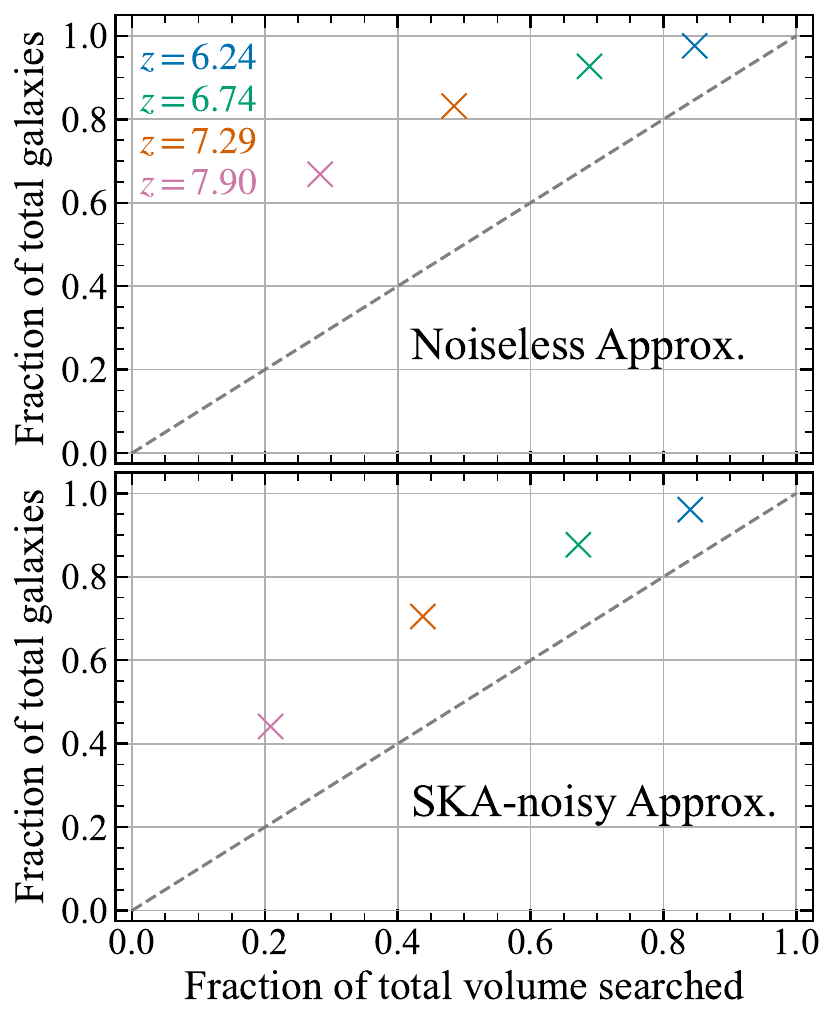}
    \caption{The mean fraction of the total galaxy population present in the recovered ionized regions of noiseless (top panel) and SKA-noisy (bottom panel) approximate light-cones at redshifts $6.24, 6.74, 7.29$ and $7.90$.}
    \label{fig:optimization_plot}
\end{figure}

\noindent The mean $\Phi_{\text{GTI}}$, $\Phi_{\text{GTN}}$, $\Phi_{\text{PI}}$, $\Phi_{\text{PN}}$, and  $\Phi_{\text{Global}}$ computed over 50 noiseless and noisy approximate light-cone galaxy catalogues are presented in Figure \ref{fig:LFs_curves}. There is a notable discrepancy between the magnitudes of $\Phi_{\text{GTN}}$ and $\Phi_{\text{GTI}}$ ($\Phi_{\text{GTI}} \gg \Phi_{\text{GTN}}$), given the vast majority of galaxies reside in ionized regions of the ground-truth maps (see Figure \ref{fig:HMFs}). While the gap between $\Phi_{\text{PN}}$ and $\Phi_{\text{PI}}$ is noticeably smaller, across both the noiseless and SKA-noisy datasets the majority of galaxies are still located in ionized regions of the recovered light-cones. This suggests that a galaxy search limited to the ionized regions of recovered light-cones will yield the largest fraction of the total galaxy population. Thus, we may optimize follow-up observations by allocating less observing time to regions with a lower probability of containing galaxies (neutral regions). To determine the redshift at which a targeted search is most efficient, we consider Figure~\ref{fig:optimization_plot}. There, we plot on the horizontal axis the fraction of the simulated volume that is labelled as ionized by our network. On the vertical axis we show the fraction of total galaxy counts that are located in the predicted ionized regions. The grey-dashed line represents the scenario where galaxies are randomly distributed in the simulation volume. Of the four redshifts considered in Figure \ref{fig:optimization_plot}, our proposed scheme of using recovered ionization maps as guides for galaxy searches is most efficient at redshift $7.90$ for the noiseless dataset and redshift $7.29$ for the SKA-noisy dataset (in the sense that these would maximize the number of galaxies found per volume searched).

\begin{figure*}
	\includegraphics[width=\textwidth]{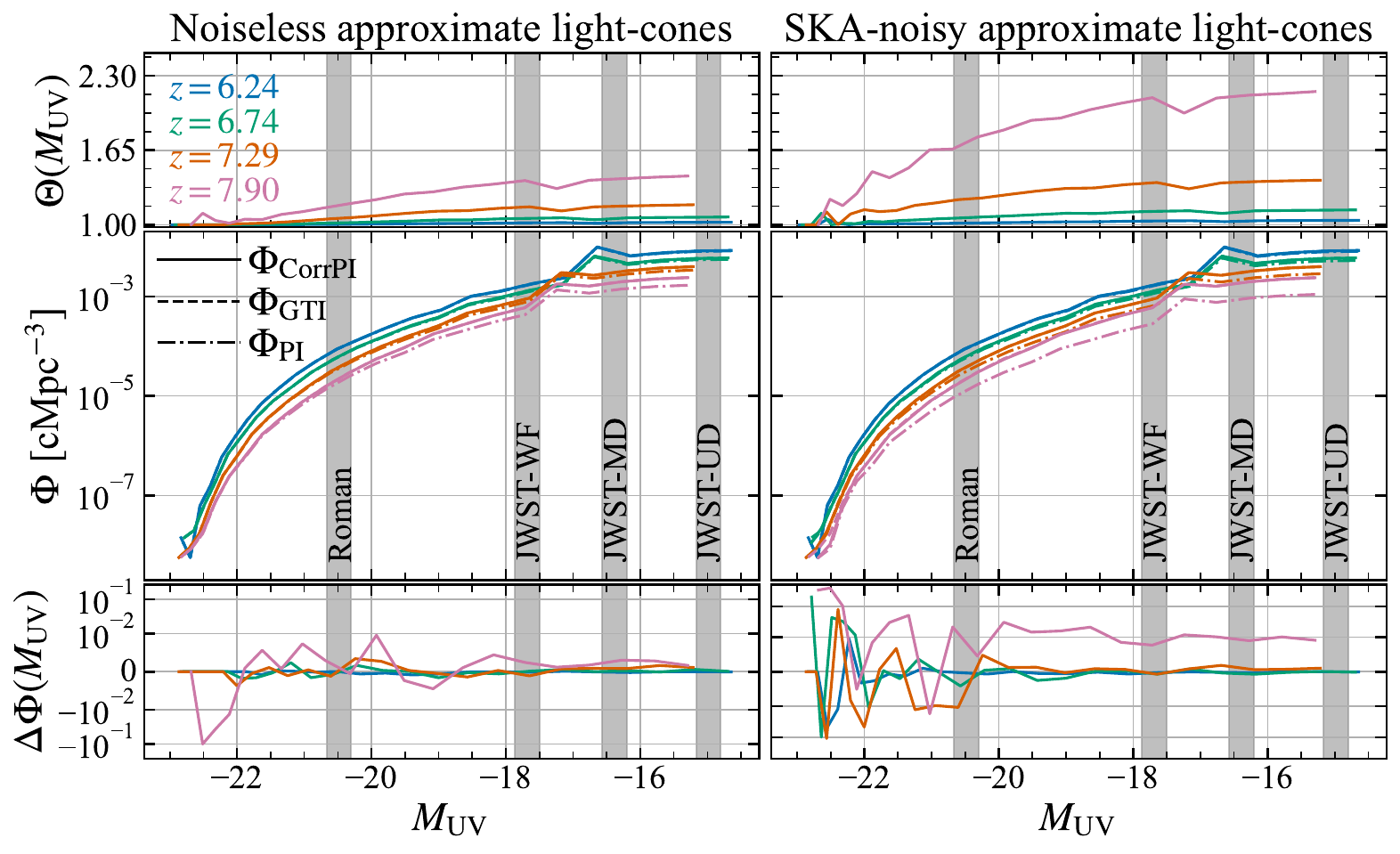}
    \caption{First row: The ratios of $\Phi_{\text{GTI}}$ to $\Phi_{\text{PI}}$ for the noiseless and SKA-noisy datasets at $z=[6.24,6.74,7.29,7.90]$. The range of limiting absolute UV-magnitudes for representative \emph{Roman}, JWST-WF, JWST-MD, and JWST-UD galaxy surveys are plotted as grey bars. Their width is determined by the minimum $M_{\text{UV}}$ that is detectable at the minimum and maximum redshifts we consider, provided the fiducial limiting apparent magnitudes of $32$, $30.6$, $29.3$, and $26.5$ for the JWST-UD, JWST-MD, JWST-WF, and \emph{Roman} galaxy surveys, respectively (\citealt{Mason2015}). All galaxies with a $M_{\text{UV}}$ less than the left edge of each survey's grey bar are luminous enough to be detected by the survey at all redshifts we consider. Second row: Noiseless and SKA-noisy GTI (dashed line) and PI (dot-dashed line) LFs from an additional set of 50 light-cone galaxy catalogues, plotted alongside the corrected PI (CorrPI, solid line) LF computed using the correction factors in the first row. Third row: The relative error between $\Phi_{\text{GTI}}$ and $\Phi_{\text{CorrPI}}$ for the noiseless and SKA-noisy datasets.}
    \label{fig:corr_fac_LFs_curves}
\end{figure*}

If one is to use the ionized regions of recovered light-cones to determine the LF of galaxies in ionized regions during the EoR, the inferred LF would follow the definition of $\Phi_{\text{PI}}$. Given we are interested in $\Phi_{\text{GTI}}$, we define an absolute magnitude-dependent correction factor $\Theta(M_{\text{UV}})$ to provide a mapping from the inferred LF to the true LF. This multiplicative correction factor is defined as
\begin{equation}
    \Theta(M_{\text{UV}}) \equiv \frac{\langle \Phi_{\text{GTI}}(M_{\text{UV}}) \rangle}{\langle \Phi_{\text{PI}}(M_{\text{UV}})\rangle},
    \label{eq:corr_fac}
\end{equation}
where $\langle \dots \rangle$ denotes an average over our ensemble of simulations. This correction factor, computed using the LFs in Figure \ref{fig:LFs_curves}, is shown in the first row of Figure \ref{fig:corr_fac_LFs_curves}. To evaluate the generality of this correction factor, we compute the mean $\Phi_{\text{GTI}}$ and $\Phi_{\text{PI}}$ over an additional set of 50 light-cone galaxy catalogues separate from those used to compute $\Theta(M_{\text{UV}})$. These are shown in the second row of Figure \ref{fig:corr_fac_LFs_curves}, alongside the corrected PI LF, $\Phi_{\text{CorrPI}}(M_{\text{UV}}) = \Theta(M_{\text{UV}}) \times \Phi_{\text{PI}}(M_{\text{UV}})$. To evaluate how well $\Phi_{\text{CorrPI}}$ agrees with $\Phi_{\text{GTI}}$, we compute the relative error 
\begin{equation}
    \Delta \Phi (M_{\text{UV}}) = \frac{\Phi_{\text{GTI}}(M_{\text{UV}})-\Phi_{\text{CorrPI}}(M_{\text{UV}})}{\Phi_{\text{GTI}}(M_{\text{UV}})}
\end{equation}
in the third row of Figure \ref{fig:corr_fac_LFs_curves}. The range of limiting absolute magnitudes of various present and upcoming galaxy surveys at the relevant redshifts are plotted as vertical bars in all subplots Figure \ref{fig:corr_fac_LFs_curves}. In this work, we consider representative JWST ultra-deep (UD), medium-deep (MD), wide-field (WF), and \emph{Roman} galaxy surveys. Observing where the different survey thresholds intersect the curves in Figure \ref{fig:corr_fac_LFs_curves} provides an indication of how significant a survey's theoretically observable galaxy population will be impacted by $\Theta(M_{\text{UV}})$. For example, given \emph{Roman} is projected to observe only the brightest galaxies at EoR redshifts, this coincides with the domain where  $1.2 \gtrapprox \Theta(M_{\text{UV}}) \gtrapprox 1.0$ and $1.7 \gtrapprox \Theta(M_{\text{UV}}) \gtrapprox 1.0$, for the noiseless and noisy light-cone galaxy catalogues, respectively. Conversely, surveys such as JWST-UD will be able to observe nearly the entire LF, requiring the application of a larger correction factor to the fainter end.

As redshift decreases, the vertical offset between $\Phi_{\text{GTI}}$ and $\Phi_{\text{PI}}$ in Figure \ref{fig:corr_fac_LFs_curves} decreases as well, resulting in curves that nearly completely overlap by redshift $6.24$. As a result, $\Theta(M_{\text{UV}}, z=6.24) \approx 1$ across nearly the entire $M_{\text{UV}}$ range. This trend is consistent across both the noiseless and SKA-noisy galaxy catalogues, and may be attributed to the overall improvement in U-Net recovery performance at lower redshifts. It is important to note that $\Theta(M_{\text{UV}}) > 1$ for all four redshifts we consider. This is due to our network ``under-ionizing": producing predicted light-cones with a greater neutral fraction than the ground-truth light-cones. Figure \ref{fig:fake_lc_preds} demonstrates this explicitly, whereby there are fewer ionized (white) regions in the predictions compared to the ground-truth. As a result, fewer galaxies are located in predicted ionized regions. This discrepancy is increasingly prevalent at higher redshifts, explaining the growing amplitude of $\Theta(M_{\text{UV}})$ in Figure \ref{fig:corr_fac_LFs_curves} as a function of redshift. The underlying $M_{\text{UV}}$-evolution of $\Theta(M_{\text{UV}})$ at higher redshifts suggests that our network more reliably recovers the ionized regions containing brighter galaxies (or more massive halos). Given our U-Net is well-suited to consistently identify the largest ionized regions, this indicates a relationship between galaxy luminosity and ionized bubble size, whereby the brightest galaxies (or most massive halos) reside in the largest ionized bubbles (although in more detailed models there can be situations where the brightest galaxies do not always reside in the centres of large ionized regions; see \citealt{Mirocha_2021}). In principle, this relationship can be leveraged to further improve the performance of our network, but we leave this possibility to future study.

The relative error between $\Phi_{\text{GTI}}$ and $\Phi_{\text{CorrPI}}$ in Figure \ref{fig:corr_fac_LFs_curves} varies considerably at lower $M_{\text{UV}}$. This may be attributed to a higher variance in the number of extremely bright galaxies (or the most massive halos) present in each simulation volume. The growth of $\Delta \Phi(M_{\text{UV}})$ with increasing redshift may also be attributed to the higher variance in recovery performance presented in Figure \ref{fig:lightcone_rec_stats}.

\section{Conclusions}\label{sec:5}

In this paper, we have successfully expanded upon the U-Net recovery framework originally presented in GH21. As an important development to the work of GH21, we extend the U-Net's capability to process 21-cm light-cones, more closely resembling the observations that will be available from interferometers such as SKA1-Low in the near future. In parallel with this, we perform hyperparameter optimization to improve the overall predictive performance of the U-Net algorithm. We demonstrate that our network is able to reliably identify the location and morphology of the largest ionized regions in foreground wedge removed 21-cm images at EoR redshifts (see Figure \ref{fig:fake_lc_preds}). Our investigations underline that the U-Net recovery retains some level of reliability even when the instrumental limitations and noise of SKA1-Low are considered, exhibiting a manageable redshift-dependent downturn in predictive performance. We detail the U-Net's redshift-dependent performance as a function of various binary classification metrics and outline the extent of the U-Net's reconstruction effort across different spatial scales using the normalized cross-power spectrum of the ground-truth and prediction light-cones (see Figures \ref{fig:coeval_rec_stats}, \ref{fig:lightcone_rec_stats}, \ref{fig:norm_CrossCorr_coeval} and \ref{fig:norm_CrossCorr_lightcones}). 

As the principal advancement of this work, we establish a connection between the U-Net recovery framework and high-redshift galaxy catalogues. In doing so, we illustrate how U-Net recovered light-cones can provide information regarding the ionization state of the IGM surrounding the galaxies that will be surveyed by current and next-generation instruments such as JWST and \emph{Roman}. We subsequently outline how the ionized regions of recovered light-cones may be used as a guide for follow-up observations of high-redshift galaxies. We additionally demonstrate how the luminosity function of galaxies located in the ionized bubbles of U-Net recovered light-cones can be corrected to recover the true LF of galaxies in ground-truth ionized regions. We provide estimates of the luminosity-dependent correction factor and evaluate the efficacy of a targeted galaxy search over a range of EoR redshifts (see Figures \ref{fig:corr_fac_LFs_curves} and \ref{fig:optimization_plot}).

In future work, comparing the distribution of ionized bubble radii in the ground-truth and recovered light-cones will provide further quantitative insight into the U-Net recovery effort. This will allow for the practical study of the relationship between galaxies (and their properties) and the radii of the ionized regions the galaxies reside in. Given this statement \textit{and} the simulation model we employ conventionally assumes an inside-out model of reionization, we acknowledge that alternative outside-in models are not considered in this work. Modifying the underlying simulation architecture to account for outside-in reionization (see \citealt{Pagano_and_Liu_2020,2021MNRAS.508.1915P}) may therefore drastically change the conclusions of our galaxy survey-related investigations (given galaxies would now preferentially reside in neutral regions). In this direction, future work may also benefit from a generalization to a variable set of astrophysical and cosmological parameters that more accurately reflects our current understanding of reionization. As such, we recognize the validity of our results in this proof-of-concept study are indeed somewhat limited to the fiducial set of parameters and processes assumed in our suite of \texttt{21cmFASTv3} simulations. Subsequent analysis may also benefit from incorporating the information present in high-redshift galaxy catalogues into the existing U-Net framework. Providing galaxy location information alongside foreground wedge-removed 21-cm images may serve to improve the overall reconstruction fidelity and is the subject of future work. The implementation of alternative machine learning-based models may also result in better reconstructions. In particular, the use of a probabilistic model (e.g. denoising U-Nets; \citealt{masipa2023emulating}) may improve the recovery of small scale power where our deterministic U-Net is currently lacking.

Coupling next-generation 21-cm interferometers with upcoming high-redshift galaxy surveys will enable further insight into how the high-redshift galaxy luminosity function varies across ionization environments during the EoR. Developing novel data analysis frameworks that both mitigate astrophysical foreground contamination \textit{and} exploit the complementarity of these two classes of observations will ultimately sharpen our understanding of the EoR and the sources that drive its evolution.

\section*{Acknowledgements}

The authors are delighted to thank Matteo Blamart, Rebecca Ceppas de Castro, Franco Del Balso, Yael Demers, Hannah Fronenberg, Adelie Gorce, Lisa McBride, Andrei Mesinger, Steven Murray, Michael Pagano, and Bobby Pascua for useful discussions. AL, JK, and JCC acknowledge support from the Trottier Space Institute, the New Frontiers in Research Fund Exploration grant program, the Canadian Institute for Advanced Research (CIFAR) Azrieli Global Scholars program, a Natural Sciences and Engineering Research Council of Canada (NSERC) Discovery Grant and a Discovery Launch Supplement, the Sloan Research Fellowship, and the William Dawson Scholarship at McGill. JM was supported by an appointment to the NASA Postdoctoral Program at the Jet Propulsion Laboratory/California Institute of Technology, administered by Oak Ridge Associated Universities under contract with NASA. 

%The Acknowledgements section is not numbered. Here you can thank helpful colleagues, acknowledge funding agencies, telescopes and facilities used etc. Try to keep it short.

%%%%%%%%%%%%%%%%%%%%%%%%%%%%%%%%%%%%%%%%%%%%%%%%%%
\section*{Data Availability}

The data underlying this article is available upon request. The 21-cm brightness temperature fields and corresponding halo catalogues can be re-generated from scratch using the publicly available $\texttt{21cmFASTv3}$ code (\citealt{Murray_2020_JOSS}). The instrumental noise and $uv$-coverage realizations can also be re-generated using the publicly available $\texttt{tools21cm}$ code (\citealt{Giri_tools21cm}). The foreground wedge-removal code, U-Net code, and the set of optimal model weights are all available upon request. 



\bibliographystyle{mnras}
\bibliography{mnras_template} % if your bibtex file is called example.bib

%\begin{thebibliography}{99}
%\end{thebibliography}

%%%%%%%%%%%%%%%%%%%%%%%%%%%%%%%%%%%%%%%%%%%%%%%%%%

%%%%%%%%%%%%%%%%% APPENDICES %%%%%%%%%%%%%%%%%%%%%

%\appendix



%%%%%%%%%%%%%%%%%%%%%%%%%%%%%%%%%%%%%%%%%%%%%%%%%%

% Don't change these lines
\bsp	% typesetting comment
\label{lastpage}
\end{document}